\documentclass[apj]{emulateapj}
\usepackage{comment}

\newcommand{\lya}        {Ly$\alpha$}
\newcommand{\llya}       {$L_{\rm Ly\alpha}$}
\newcommand{\Aiso}       {$A_{\rm iso}$ }
\newcommand{\unitcgssb}  {ergs\,s$^{-1}$\,cm$^{-2}$\,arcsec$^{-2}$}

\newcommand{\unitcgslum} {ergs\,s$^{-1}$}

\newcommand{\nb}         {{\sl NB}}

\newcommand{\U}          {{\sl U}}
\newcommand{\B}          {{\sl B}}
\newcommand{\UU}         {{\sl U }}
\newcommand{\BB}         {{\sl B }}
\newcommand{\CIV}        {\hbox{{\rm C}\kern 0.1em{\sc iv}}}

\newcommand\avg[1]       {\langle#1\rangle}
\newcommand{\X}          {\phn{ }}

\slugcomment{ApJ in press}

\shorttitle{Statistics of Ly$\alpha$ Blobs}
\shortauthors{Yang et al.}

\begin{document}

\title{Strong Field-to-Field Variation of L\lowercase{y$\alpha$}
Nebulae Populations at \lowercase{$z \simeq 2.3$}}

\author{
Yujin Yang\altaffilmark{1,2}, 
Ann Zabludoff\altaffilmark{2}, 
Daniel Eisenstein\altaffilmark{2}, 
Romeel Dav\'e\altaffilmark{2}}

\altaffiltext{1}{Max-Planck-Institut f\"ur Astronomie, K\"onigstuhl 17,
D-69117, Heidelberg, Germany. yyang@mpia.de}
\altaffiltext{2}{Steward Observatory, University of Arizona, 933 North Cherry Avenue, Tucson AZ 85721}



\begin{abstract}

Understanding the nature of distant \lya\  nebulae, {\it aka} ``blobs,''
and connecting them to their present-day descendants requires constraining
their number density, clustering, and large-scale environment.  To measure
these basic quantities, we conduct a deep narrowband imaging survey in
four different fields, Chandra Deep Field South (CDFS), Chandra Deep
Field North (CDFN), and two COSMOS subfields, for a total survey area of
1.2\,deg$^2$.  We discover 25 blobs at $z =2.3$ with \lya\ luminosities
of \llya = 0.7--8 $\times$ $10^{43}$ \unitcgslum\ and isophotal areas
of \Aiso = 10 -- 60\,\sq\arcsec. The transition from compact \lya\
emitters (\Aiso $\sim$\,a few \sq\arcsec) to extended \lya\ blobs (\Aiso
$>$ 10\,\sq\arcsec) is continuous, suggesting a single family perhaps
governed by similar emission mechanisms.
Surprisingly, most blobs (16/25) are in one survey field, the CDFS.  The
six brightest, largest blobs with \llya\,$\gtrsim$ 1.5$\times$10$^{43}$
\unitcgslum\ and \Aiso $>$ 16\,\sq\arcsec\ lie {\it only} in the
CDFS.  These large, bright blobs have a field-to-field variance of
$\sigma_v$ $\gtrsim$ 1.5 (150\%) about their number density $n$ $\sim$
$1.0^{+1.8}_{-0.6}$$\times$ $10^{-5}$ Mpc$^{-3}$.  This variance is large,
significantly higher than that of unresolved \lya\ emitters ($\sigma_v$
$\sim$ 0.3 or 30\%), and can adversely affect comparisons of blob number
densities and luminosity functions among different surveys.  Our deep,
blind survey allows us to construct a reliable blob luminosity function.
We compare the statistics of our blobs with dark matter halos in a 1
$h^{-1}$ Gpc cosmological N-body simulation.  At $z=2.3$, $n$ implies that
each bright, large blob could occupy a halo of $M_{\rm halo}$ $\gtrsim$
$10^{13}$\,$M_{\sun}$ if most halos have detectable blobs.  The predicted
variance in $n$ is consistent with that observed and corresponds to a bias
of $\sim$7.  Blob halos lie at the high end of the halo mass distribution
at $z=2.3$ and are likely to evolve into the $\sim$10$^{14}$ $M_{\sun}$
halos typical of galaxy clusters today.  On larger scales of $\sim 10$
co-moving Mpc, blobs cluster where compact \lya\ emitters do, indicating
that blobs lie in coherent, highly overdense structures.


\end{abstract}


\section{Introduction}

\lya\ nebulae, or ``blobs,'' are extended sources at $z$ $\sim$ 2--6
with typical \lya\ sizes of $\gtrsim$\,5\arcsec\ ($\gtrsim$50\,kpc) and
line luminosities of $L_{\rm{Ly\alpha}}\gtrsim10^{43}$ \unitcgslum\
\cite[e.g.,][]{Keel99, Steidel00, Francis01, Matsuda04, Dey05,
Smith&Jarvis07, Hennawi09, Prescott09, Yang09}.  Because the large
spatial extent of their \lya-emitting gas implies an interaction
between the surrounding intergalactic medium and any embedded galaxies,
blobs may signal an important phase of galaxy formation in the early
universe, including cold gas accretion \citep{Haiman00, Fardal01,
Yang06, Dijkstra&Loeb09}, galactic-scale feedback due to stellar winds
\citep{Taniguchi&Shioya00}, or intense radiative feedback from AGN
\citep{Geach09}.
Despite the importance of blobs and the controversy regarding their
origins, even basic properties such as their number density, clustering,
and large-scale environment are poorly constrained.

To understand into what these mysterious objects will evolve in the
present day universe, measuring their statistics is critical due to the
direct connection of number density and field-to-field variance to halo
mass in $\Lambda$CDM cosmology.
Currently, the halo mass of blobs is unknown.  Using the spatial extent
and line-width of the \lya\ line, \citet{Matsuda06} estimate dynamical
masses of 0.5 -- 20 $\times$ 10$^{12}$ $M_{\sun}$ if the extended \lya\
emission is from gravitationally bound gas clouds and the resonant
scattering of \lya\ can be ignored.


The clustering of blobs has not been measured directly in past work,
yet there are hints  that it is strong.  After surveying over $\sim$4.8
deg$^2$ in the NOAO Deep-Wide Bo\"otes field \citep{Jannuzi&Dey99},
\citet{Yang09} discover just four bright blobs, yet two of them
lie within only $\sim$70\arcsec\ of each other \cite[see also the
discovery of a \lya\ blob near a radio-loud \lya\ halo by][]{Matsuda09}.
Some, but not all \cite[e.g.,][]{Gronwall07,Nilsson09}, narrowband
surveys targeting {\it compact} \lya\ emitters also detect blobs at similar
redshifts.  For example, while following up the two bright blobs found
by \citet{Steidel00} in the SSA22 field, \citet{Matsuda04} discover
a spectacular clustering of 35 blobs, in which the two brightest
lie at the intersection of filaments traced by compact \lya\ emitters
\citep{Matsuda05}. \citet{Palunas04} discover two\footnotemark\ additional
\lya\ blobs in the J2143-4423 region defined by an overdensity of compact
\lya\ emitters \citep{Francis01}.
\footnotetext{One of these blob candidates has now been identified as
a low-z interloper (J.~Colbert priv.~comm.)}
To measure directly the number density and clustering of \lya\ blobs,
and thus to constrain their halo mass, requires a large volume survey,
particularly over different sight lines, to account for any field-to-field
variations.  Furthermore, one must apply uniform selection criteria for
identifying blobs over the entire survey volume.

To acquire a large, unbiased sample of \lya\ blobs at $z=2.3$, we have
pursued two complimentary narrow-band imaging surveys.  The shallow,
but wide sky coverage, survey using the Steward Observatory Bok 2.3m +
90Prime imager targets rare, luminous \lya\ blobs \cite[\llya\ $\gtrsim$
$2\times10^{43\ }$ \unitcgslum;][]{Yang09}.  In this paper, we report the
first results from our deeper, but smaller sky coverage, survey with the
NOAO 4m telescopes and MOSAIC imagers that targets presumably more
common intermediate size and luminosity \lya\ blobs like those
discovered by \citet{Matsuda04}.
Using blob statistics from the four different
30\arcmin\,$\times$\,30\arcmin\ survey fields, in the Chandra Deep
Field-South \cite[CDFS;][]{Brandt01}, Chandra Deep Field-North
\cite[CDFN;][]{Giacconi02}, and two regions of the Cosmic Evolution
Survey \cite[COSMOS;][]{Scoville07,Koekemoer07}, we determine the
field-to-field variation in the blob number density.  We then use a large
volume cosmological N-body simulation to constrain their host halo masses
for the first time.

In \S\ref{sec:observation}, we describe our narrow-band imaging survey and
the data reduction procedures. \S\ref{sec:sample_selection} describes
the selection of the \lya\ blob sample.  In \S\ref{sec:results},
we present the \lya\ blob candidates (\S\ref{sec:sample}), compare
their sizes and luminosities to those of compact \lya\ sources
(\S\ref{sec:sequence}), constrain their field-to-field variation and halo
masses (\S\ref{sec:cosmic_variance}), and characterize their large-scale
environment using the compact \lya\ emitters (\S\ref{sec:lss}).
In \S\ref{sec:conclusion}, we discuss our conclusions.  Throughout this
paper, we adopt the cosmological parameters $H_0$ = 100\,$h$\,${\rm
km\,s^{-1}\ Mpc^{-1}}$, $h = 0.7$, $\Omega_{\rm M}=0.3$, and
$\Omega_{\Lambda}=0.7$. All magnitudes are in the AB system \citep{Oke74}.


\section{Observations and Data Reduction}
\label{sec:observation}


Using the MOSAIC--I and II CCD imagers on the KPNO Mayall and the CTIO
Blanco 4m telescopes, we obtain deep narrowband images with a custom
narrowband filter (hereafter \nb403 or \nb). This narrowband filter
has a central wavelength of $\lambda_c \approx 4030$\AA, designed for
selecting \lya-emitting sources at $z\approx2.3$.  Its band-width of
$\Delta\lambda_{\rm FWHM} \approx 45$\AA\ provides a line-of-sight
depth of $\Delta z \simeq 0.037$, corresponding to 46.8\,Mpc at $z=2.3$
in the comoving frame.  The CDFS, CDFN, and COSMOS survey fields have
extensive ancillary data sets, including the deepest X-ray images for
robust identification of AGN.

We conducted the narrowband imaging observations over nine photometric
nights between January 2007 and February 2009.  The MOSAIC I and II
cameras have eight 2k\,$\times$\,4k CCDs with a pixel scale of 0\farcs27
pixel$^{-1}$, leading to a sky coverage of 37\arcmin\,$\times$\,37\arcmin.
We obtain deep narrowband images for four different pointings: CDFS,
CDFN, and two COSMOS subfields (hereafter COSMOS1 and COSMOS2).  The total
exposure time ranges from 7.2 to 10 hr, which consists of individual
20 or 30 minute exposures with a standard dither pattern to fill in the
gaps between the eight chips.
The seeing ranges from 1\farcs0 to 1\farcs3 depending on the fields.
Table \ref{tab:narrowband} summarizes our narrowband observations,
including the central coordinates of the survey fields, total exposure
times, survey areas, seeing, and survey depths.

\begin{figure}[!t]
\epsscale{1.2}
\plotone{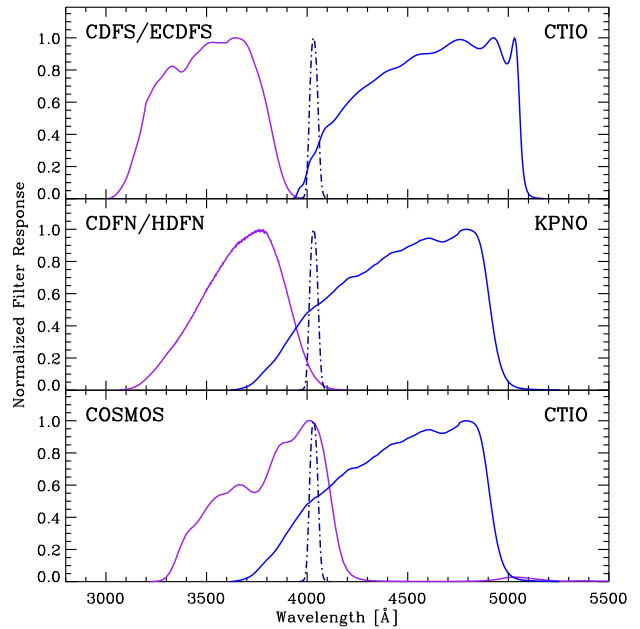} 
\caption{
Filter response profiles for the \nb403 narrowband ({\it dot-dashed line})
and broadband (\UU and \B) filters ({\it solid lines}) used in this study.
The profiles  are normalized to a maximum throughput of 1 and include
the transmission of the atmosphere, telescope, camera optics, filter, and detector.
({\it Top}) ESO 2.2m \UU and \BB band filters from MUSYC \citep{Gawiser06a,Gawiser06b}.
({\it Middle}) KPNO \UU and Subaru \BB filters from the Hawaii-HDF survey \citep{Capak04}.
({\it Bottom}) CFHT {\sl u}$^{*}$ and Subaru \BB filters from COSMOS \citep{Capak07}.
}
\label{fig:filter_response}
\end{figure}

To identify line emission objects requires that we subtract the continuum
emission underlying the \nb403 bandpass. We estimate the continuum using
existing, deep broadband (\UU and \B) images.
For the CDFS, we use optical images from the Extended Chandra Deep
Field-South dataset from the Multiwavelength Survey by Yale-Chile
\cite[MUSYC;][]{Gawiser06a,Gawiser06b}\footnotemark.
\footnotetext{The original data were taken with the ESO MPG 2.2\,m
and Wide Field Imager (WFI) by the ESO Deep Public Survey and COMBO-17
teams \citep{Arnouts01,Wolf04,Erben05,Hildebrandt06}.  In this paper,
we use the data products delivered by the MUSYC team.}
For the CDFN, we use the imaging products from the Hawaii Hubble Deep
Field North survey \citep{Capak04}.  For COSMOS1 and COSMOS2, we use 25
image tiles for each subfield\footnotemark\ produced by \citet{Capak07}.
For CDFS and CDFN, the broadband images have smaller sky coverage
than our narrowband images, limiting our final survey areas (Table
\ref{tab:narrowband}).  Table \ref{tab:broadband} lists details of the
broadband images, including the filter properties, survey depth, seeing,
and the instruments used.  We also show the narrow and broadband filter
transmission curves for all fields in Figure \ref{fig:filter_response}.
\footnotetext{The tile numbers for the lower-left and upper-right corners
of COSMOS1 and COSMOS2 are (17, 69) and (65, 117), respectively.}


We reduce the narrowband images with the IRAF {\sl mscred} mosaic data
reduction package \citep{Valdes98} following the procedures of the NOAO
Deep Wide Field Survey team \citep{Jannuzi&Dey99}. The data are corrected
for crosstalk between amplifiers and bias-subtracted.  For flat-fielding,
we use dome flats together with night-sky flats, which are median-combined
from unregistered object frames each night. Satellite trails, CCD edges,
bad pixels, and saturated pixels are masked.  The astrometry is calibrated
with the USNO-B1.0 catalog \citep{Monet03} using the IRAF {\sl ccmap}
task.
The individual images are transformed to have the same pixel scales and
world coordinate systems (WCS) as the reference broadband images.  For the
COSMOS broadband images with finer pixel scale (0\farcs15 pixel$^{-1}$),
we resample them with the coarse MOSAIC--II pixel scale (0\farcs27
pixel$^{-1}$) to make the reference images.  Finally, the projected
images are scaled using common stars in each frame and stacked to remove
cosmic rays.  For flux calibration, we employ the 3--5 spectrophotometric
standard stars observed each night to derive extinction coefficients
and zero points for the \nb403 magnitudes. Typical uncertainties in the
derived zero-points are 0.02\,--\,0.04 mag.  These uncertainties are
added to the photometry errors in quadrature.

In addition to the standard reduction procedures described above, we
pay special attention to the ``crosstalk'' that occurs between CCDs (or
amplifiers) sharing the readout electronics.  The net effect of crosstalk
is that a very bright or saturated source on one chip (or amplifier)
produces echos or ghost images on the paired chips with proportional
intensities, typically 0.02--0.2\%, up to several counts.  One can
determine these proportionalities, known as crosstalk coefficients,
using the science image itself, and thus remove echo images.  In most
circumstances, this crosstalk correction works well and does not affect
the detection and photometry of bright sources. Furthermore, dithered
exposures often average out the ghost images in the individual frames
when stacked to make the final products.

However, residual ghost images from imperfect correction can be easily
confused with low surface brightness objects.  In particular, for CTIO
MOSAIC--II, the crosstalk images of a source at pixel ($x$, $y$) on one
chip appears at the same pixel coordinates on the other chips (victim
chips). Therefore, ghost images always appear at the same location
relative to real sources in the dithered exposures, and they will be
co-added in the final combined images, leading to false detections
mimicking diffuse extended emission.


\begin{figure}[!t]
\epsscale{1.15}
\plotone{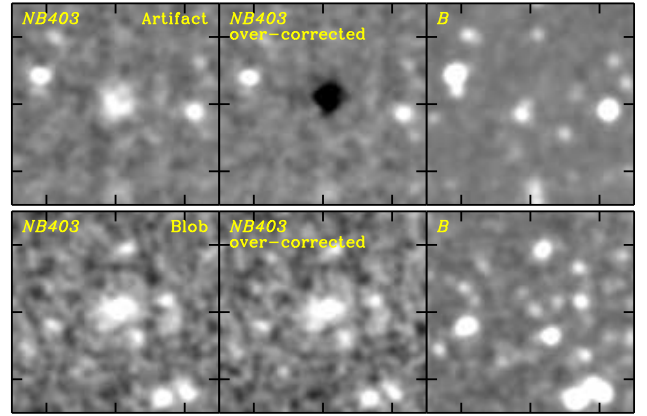} 
\caption{
Distinguishing between a false detection arising from an imperfect
crosstalk correction of the CTIO MOSAIC--II image and a true extended
\lya\ source.
({\it From left to right}) Original \nb403 images, \nb403 images
over-corrected with crosstalk coefficients that are two times too large,
and {\sl B} band images. The ticks represent 10\arcsec\ intervals.
Note that the residual due to the imperfect crosstalk correction arising
from the standard reduction ({\it top left panel}) resembles an \lya\
source that is spatially more extended than its continuum counterpart ({\it
top right panel}).  This artifact appears as a negative mirror image
in the over-corrected image ({\it top middle panel}), while the real
object is not affected by the crosstalk over-correction ({\it bottom
middle panel}).
By comparing the original and over-corrected images, we are able to reject
false detections, which is critical given the small number statistics
of the blob survey.
}
\label{fig:crosstalk_correction}
\end{figure}


To weed out false detections, we repeat the entire reduction procedure
doubling the crosstalk coefficients so that the ghost images are
over-corrected and appear as negative counts.  If a blob candidate
is indeed an artifact arising from imperfect correction, it will
show up as a negative image in the final combined image.  Figure
\ref{fig:crosstalk_correction} shows an example in which one artifact
image appears as a negative mirror image, whereas a real object is not
affected by this crosstalk over-correction.  Note that the artifact is
as bright as the real blob candidate, demonstrating that special care is
required to reject false detections for extremely low-surface brightness
objects like \lya\ blobs.

\section{Selection of \lya\ Blob Candidates}

\label{sec:sample_selection}

To find \lya\ blob candidates, we construct photometric catalogs in
the \nb403 narrowband and two broadbands (\UU and \B) using SExtractor
\citep{Bertin&Arnouts96}.  First, we make ``detection'' images (\nb+\U+\B)
by adding the \nb403 and broadband images after scaling them according
to their signal-to-noise ratios (S/N).  After identifying sources in the
``detection'' images that have least 2 pixels that are 1.5$\sigma$ above
the local sky,  we run SExtractor in double-image mode on the \nb403,
\U, and \BB images using these detection images. In other words, we first
find the sources in the ``detection'' images and then obtain photometry
at their position in the \nb403, \U, and \BB images to construct three
separate (one narrow- and two broadband) catalogs.  We adopt Kron-like
elliptical aperture magnitudes (i.e., {\tt MAG\_AUTO} in SExtractor)
to derive photometric properties.  Our use of the ``detection'' images
ensures that 1) all the sources detected in either \nb\ or the broadbands
are included in our catalog and 2) the elliptical apertures of the more
extended sources in the (\nb+\U+\B) image are large enough to include
all the light from both the \nb\ and broadband images.  Note that this
choice of photometric aperture is different from the classical \lya\
emitter searches that adopt a small circular aperture (a few $\times$
FWHM of seeing) to detect fainter sources.

The selection of \lya\ blob candidates from the \nb\ and broadband
photometric catalogs consists of two steps: 1) selection for line
(hopefully, \lya) emitting objects with large line equivalent widths and
2) selection for spatially extended objects with a larger angular extent
in line emission than in the broadbands.  In other words, we define a
``blob'' as an object whose \lya\ emission above a certain surface
brightness threshold is more extended than its stellar continuum, thus
representing light from the intergalactic medium.

\begin{figure}[!t]
\epsscale{1.15}
\plotone{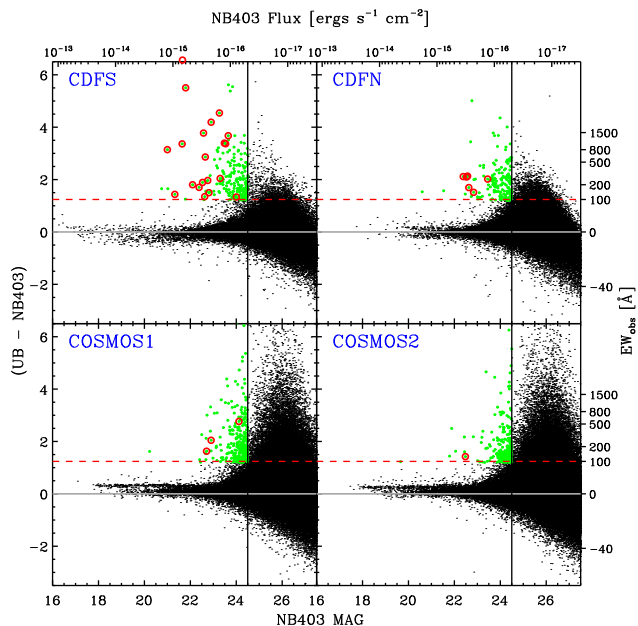}
\caption{
Color -- magnitude ({\sl UB}$-$\nb403) versus \nb403 diagram for all the
sources detected in our four survey fields: CDFS, CDFN, and two COSMOS
subfields.  Here {\sl UB} represents the AB magnitude of the average
continuum flux density within the \nb403 band estimated from the $U$ and
$B$ band images.  Right and top axes show the corresponding equivalent
widths in the observed frame and the \nb\ fluxes, respectively.  We select
line-emission objects with the criteria \nb\, $<$ 24.5 (solid line) and
({\sl UB}$-$\nb403) $>$ 1.2 (dashed line; EW$_{\rm obs}$ $>$ 100\AA).
Open circles represent the final \lya\ blob candidates obtained from
Fig.~\ref{fig:size_luminosity}. There are more open circles than final
blob candidates in the CDFS panel because a few blob candidates include
multiple sources.
}
\label{fig:color_mag}
\end{figure}

First, we choose candidates by requiring that they are detected above the
completeness limits of the \nb403 images (\nb403 $\lesssim$ 24.5 mag).
All candidates must have observed-frame equivalent widths larger than
100\AA\ (EW$_{\rm rest}$ $>$ 30.3\AA), corresponding to ($U\!B$ $-$
$N\!B$) $>$ 1.2, where {\sl UB} represents the AB magnitude of the
average continuum flux density within the \nb403 band ($f^{N\!B}_{cont}$)
estimated from the \U\ and \B\ band images, $U\!B$ $\equiv$
$-2.5 \log(f^{N\!B}_{cont}) - 48.60$.
We estimate $f^{N\!B}_{cont}$ and the line flux ($F_{line}$) of these objects
using the following relations:
\begin{eqnarray}
\label{eq:continnum-subtraction}
f^{B}_{cont}    &=& \frac{F_B - \epsilon_{B} F_{N\!B}}{\Delta\lambda_B - \Delta\lambda_{N\!B}} \\
\nonumber
f^{U}_{cont}    &=& \frac{F_U - \epsilon_{U} F_{N\!B}}{\Delta\lambda_U - \Delta\lambda_{N\!B}} \\
\nonumber
f^{N\!B}_{cont} &=& \frac{1}{\lambda_B-\lambda_U} 
                    \left[(\lambda_B - \lambda_{N\!B}) f^{B}_{cont} +
                          (\lambda_{N\!B} - \lambda_U) f^{U}_{cont}\right] \\
\nonumber
F_{line}           &=& F_{N\!B}  - f^{N\!B}_{cont} \Delta\lambda_{N\!B}, 
\end{eqnarray}
%
where $F_U$, $F_B$ and $F_{N\!B}$ are the total flux in each filter
derived from the \U, \B, and \nb403 magnitudes, respectively.
$\Delta\lambda_U$, $\Delta\lambda_B$, and $\Delta\lambda_{N\!B}$
represent the band-widths of the \U, \BB and \nb\ filters, respectively.
$f^{U}_{cont}$ and $f^{B}_{cont}$ represent the average flux density
of galaxy continuum within the \U\ and \B\ bands, respectively.
$\epsilon_{B}$ and $\epsilon_{U}$ are the correction factors that
we use to remove the \nb\ light from each broadband when estimating
$f^{U}_{cont}$ and $f^{B}_{cont}$.  Figure \ref{fig:color_mag} shows the
({\sl UB} $-$ \nb) color as a function of \nb\ magnitude for all objects
detected in either the narrow or broad bands within our survey area.


In CDFS, COSMOS1, and COSMOS2, bright (18 $<$ \nb\ $<$ 22) sources,
mostly stars and nearby galaxies, have an average ($U\!B - N\!B$)
color that is consistent with zero ($-0.11$$\pm$0.18, 0.13$\pm$0.17,
and 0.12$\pm$0.12, respectively).  In the CDFN, however, they have
an average color of ($U\!B - N\!B$) $\sim$ $-0.24\pm0.14$ mag, which
suggests absorption features in our narrowband that are very unlikely.
Because this $-0.24$ mag offset disappears when we adopt continuum
measurements (in {\sl u and g}) from the Sloan Digital Sky Survey,
we apply a small correction (0.2 mag) to the CDFN broadband magnitudes
\citep{Capak04} such that $\langle U\!B - N\!B \rangle$ $\simeq$ 0 for
the bright sources.  Although we cannot explain the apparent problem
with the CDFN broadband photometry in \citet{Capak04}, the selection of
our blob sample is unaffected by the applied correction.

At our survey redshift, the only possible interlopers are nearby
[\ion{O}{2}] $\lambda$3727 emitters at $z\approx0.08$.  However, such
objects rarely have equivalent widths larger than 100\AA\ in the rest
frame \citep{Hogg98}.  Therefore, we expect that the contamination of our
$z=2.3$ \lya\ source catalog by nearby star forming galaxies is minimal.

\begin{figure}[!t]
\epsscale{1.15}
\plotone{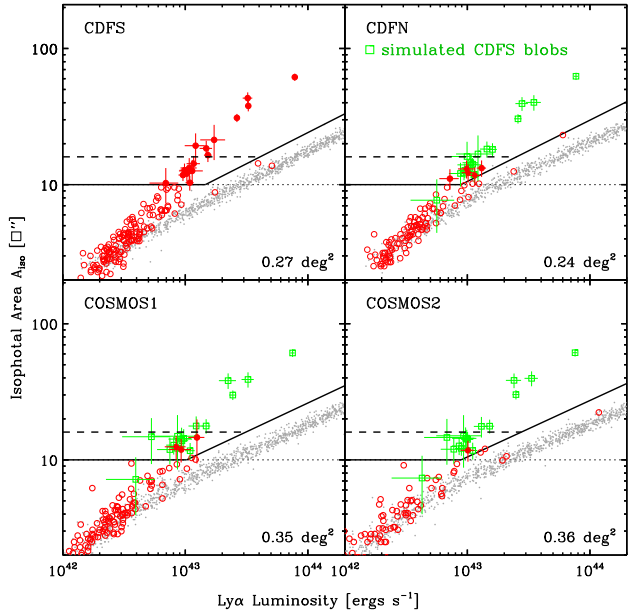}
\caption{
Distribution of isophotal areas ($A_{\rm iso}$) and \lya\ luminosities
of \lya-emitting sources (red circles) in the CDFS, CDFN, and two COSMOS
subfields.
Thick solid lines represent the selection criteria for the final
\lya\ blob candidates (filled circles): \Aiso $>$ 10\,\sq\arcsec\ and
distinguishable from point sources, i.e., deviating from the simulated
point source locus of small grey dots by $> 4\sigma$.  Simulated point
sources mix with extended \lya-emitting sources below our selection limits
in \Aiso and $L_{\rm Ly\alpha}$, making it difficult to distinguish blobs
from point sources there.  The horizontal dashed line represents the
division (\Aiso = 16\,\sq\arcsec) between the brightest, largest blobs and
the other blob candidates.   The former are discovered only in the CDFS.
The green squares in the CDFN, COSMOS1, and COSMOS2 panels represent
the \Aiso and $L_{\rm Ly\alpha}$ that the 16 CDFS blobs would have if
they were observed at the seeing and depth of each of the other fields
(see \S\ref{sec:cosmic_variance}).
The six brightest, largest blobs (\llya\,$\gtrsim$ 1.5$\times$10$^{43}$
\unitcgslum\ and \Aiso $>$ 16\,\sq\arcsec) in the CDFS should have been
detected in the other fields.  Therefore, we conclude that the observed
strong field-to-field variation is real and not due to observational
biases (see \S\ref{sec:cosmic_variance}).
}
\label{fig:size_luminosity}
\end{figure}

Second, we identify those line-emission selected objects that are
more spatially extended in \lya\ than their continuum counterparts
(Fig.~\ref{fig:size_luminosity}). This selection definition is the same
as that adopted by \citet{Matsuda04} and somewhat different than that of
\citet{Saito06}, who select spatially extended objects by requiring the
FWHM in their intermediate-band to be larger than that in the broadband
or continuum image \cite[see also][]{Nilsson09}.

We measure the spatial extent of the \lya\ emission in the
continuum-subtracted images. After registering the \nb403 and broadband
images (\UU and \B) at the sub-pixel level and matching their seeing,
we construct continuum-subtracted \nb403 images by applying the relations
in Eq.\,(\ref{eq:continnum-subtraction}) in 2-D.  We measure the isophotal
area of the emission region by running SExtractor with a threshold of $5.5
\times 10^{-18}$ \unitcgssb, which corresponds to 3$\sigma$, 1.8$\sigma$,
2.2$\sigma$, and 2.1$\sigma$ above the local sky in the CDFS, CDFN,
COSMOS1, and COSMOS2 fields, respectively.  This measurement threshold
is $\sim$ 2.5$\times$ higher than that adopted by \citet{Matsuda04}.
However, because our survey redshift ($z=2.3$) is lower than theirs
($z=3.1$), we gain a factor of $\sim$ 2.4 in surface brightness, thus
achieving equivalent surface brightness (8.6$\times$10$^5$ $L_{\sun}$
kpc$^{-2}$ compared to their 8.2$\times$10$^5$ $L_{\sun}$ kpc$^{-2}$).
We also estimate local sky background using a fairly large background
mesh size ($\sim$ 60\arcsec$\times$60\arcsec), so as not to mistakenly
subtract the extended \lya\ emission as a local background.

Measuring the size of a low surface brightness feature is always subject
to the noise and filtering.  After testing the various smoothing filters
in SExtractor, we adopt a 5$\times$5 pixel$^2$ Gaussian kernel with FWHM =
2 pixels to compromise between signal-to-noise (S/N) and over-smoothing.
We choose this truncated smoothing kernel such that it can enhance S/N in
the low surface brightness wings while not spreading the bright core into
the outer part.  We note that the measured isophotal area depends strongly
on the choice of the filters and recommend specifying the smoothing
kernel and detection threshold when reporting blob size to allow a direct
comparison between different samples.  Appendix \ref{sec:blob_comparison}
shows what sizes and fluxes the 35 blobs from \citet{Matsuda04} would have
if those blobs were measured following our procedures.  In our survey,
the \citet{Matsuda04} blobs would be smaller, and the smallest blobs would
have an isophotal area of \Aiso $\sim$ 10\,\sq\arcsec.  Therefore, we
adopt this value as the size lower-bound when selecting our blob sample.

One of the potential problems in detecting a blob is contamination by
point sources.  To quantify this effect, we place artificial point-sources
with a range of luminosities ($L_{\rm Ly\alpha}$ = 10$^{42}$ -- 10$^{45}$
\unitcgslum) into the sky regions and measure their sizes and fluxes
in the same manner as for the extended sources.  We then determine the
isophotal area versus luminosity limits above which extended and point
sources can be differentiated.


Figure \ref{fig:size_luminosity} shows the distribution of the angular
sizes and line luminosities of the \lya-emitting blob candidates assuming
that they are all at $z=2.3$.  The open and filled circles represent
the line-emitting objects selected using our line-emission criteria
(Fig.~\ref{fig:color_mag}), and the gray dots show the relation between
size and brightness for the artificial point sources.  To choose the
final blob candidates, we select objects with isophotal areas larger
than 10\,\sq\arcsec\ that lie more than 4$\sigma$ above the $L_{\rm
Ly\alpha}$--\Aiso relation defined by the point sources.  Below these
limits, extended and point sources mix, and the sizes of blobs cannot
be measured reliably, as explained below.

Because the chosen isophotal threshold is comparable to the rms sky noise
(1.8$\sigma$ -- 3$\sigma$ depending on the field), we test how reliably
we can measure the spatial extent of the blob candidates.  We cut a small
(101$\times$101 pixel$^2$) section around each candidate from the line-only
image, filter it with a smoothing kernel,\footnotemark\ place each postage
stamp into $\sim$ 1000 empty sky regions, and extract the sources with
SExtractor in the same way as for the real data.  Then we check how often
and accurately the blob sizes are recovered from these simulated images.
The recovery fraction ($f_{\rm recv}$) represents how frequently
an artificial blob is recovered with a size larger than 10\,\sq
\arcsec. The size error becomes comparable to the measured size below
our sample selection limits, and $f_{\rm recv}$ drops from $\gtrsim$90\%
to $\sim$50\% at the selection boundary for the brightest blob candidates.

Our ``blob'' definition requires their \lya\ emission to be more
spatially extended than their stellar continuum emission.  Thus,
their extended light represents the intergalactic medium instead of the
galaxies.  Our choice of size cut is not due to any discontinuity, and
is limited only by the ground-based seeing.  Therefore, higher spatial
resolution images would likely detect fainter or more compact blobs.
Stacked images of LBGs (Lyman Break Galaxies) do reveal faint extended
emission \citep{Hayashino04}.

\footnotetext{We use a 5$\times$5 pixel$^2$ convolution mask with FWHM =
2 pixel.}

\section{Results}
\label{sec:results}

\subsection{Discovery of \lya\ Blobs}
\label{sec:sample}


\begin{figure}[!t]
\epsscale{1.2}
\plotone{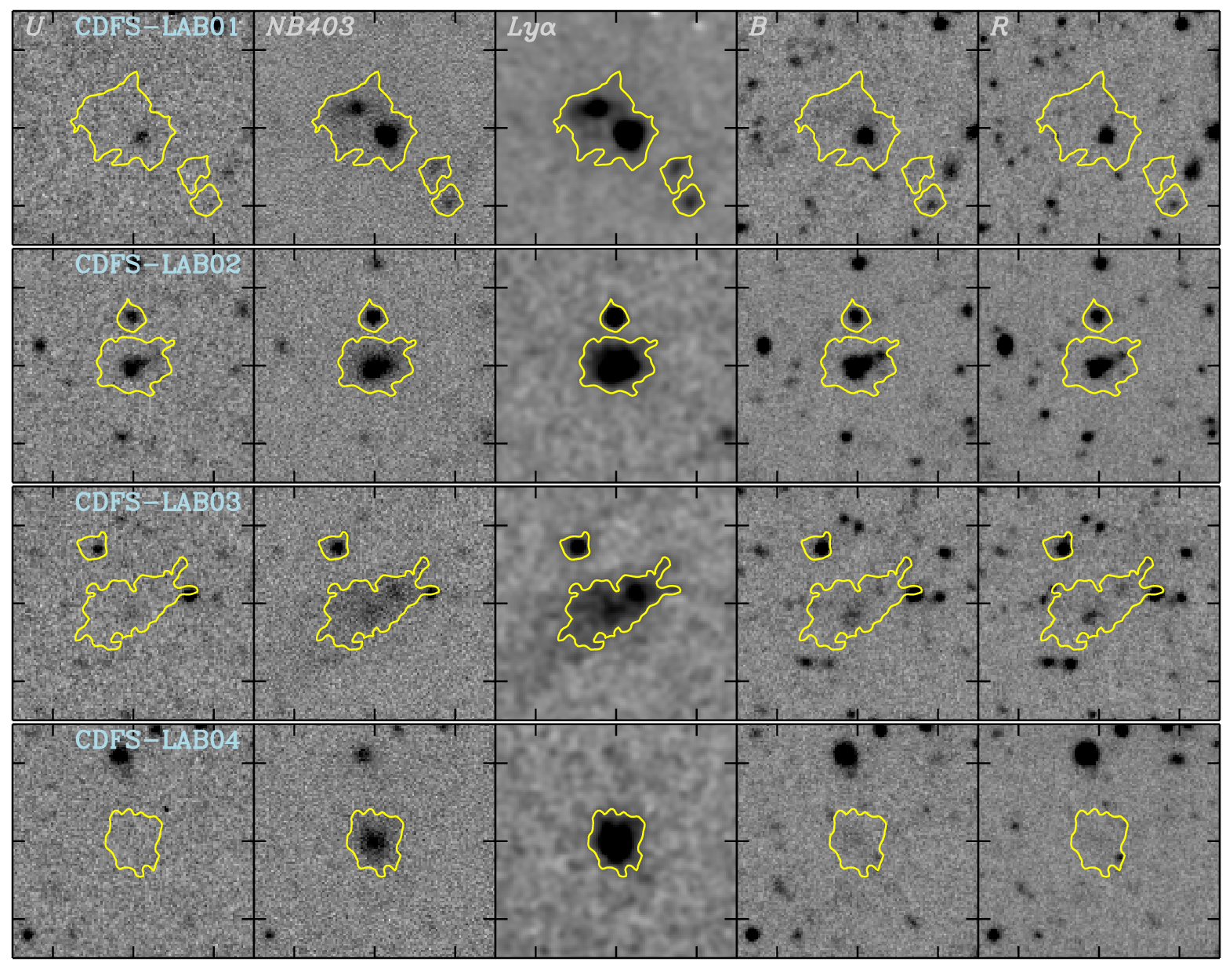}
\caption{
Cut-out images of the 16 \lya\ blobs that we discover in the Extended
CDFS.  Images from left to right:  \U, \nb403, continuum-subtracted \lya\
line ($\lambda_{\rm c}$ $\simeq$ 4030\AA), \B, and {\sl R} band.  The
ticks are spaced in 10\arcsec\ intervals.  The overlayed (yellow) contours
represent a \lya\ surface brightness of 4$\times$10$^{-18}$ \unitcgssb,
our 2.2$\sigma$ detection limit in the narrowband images. In each case,
the \lya\ emission is more extended than the broadband counterpart.
All five candidates that we have now observed spectroscopically are
confirmed as blobs at $z=2.3$ (Yang et al., in prep.).
}
\label{fig:blob_image_cdfs}
\end{figure}


\begin{figure}[!t]
\addtocounter{figure}{-1}
\epsscale{1.2}
\plotone{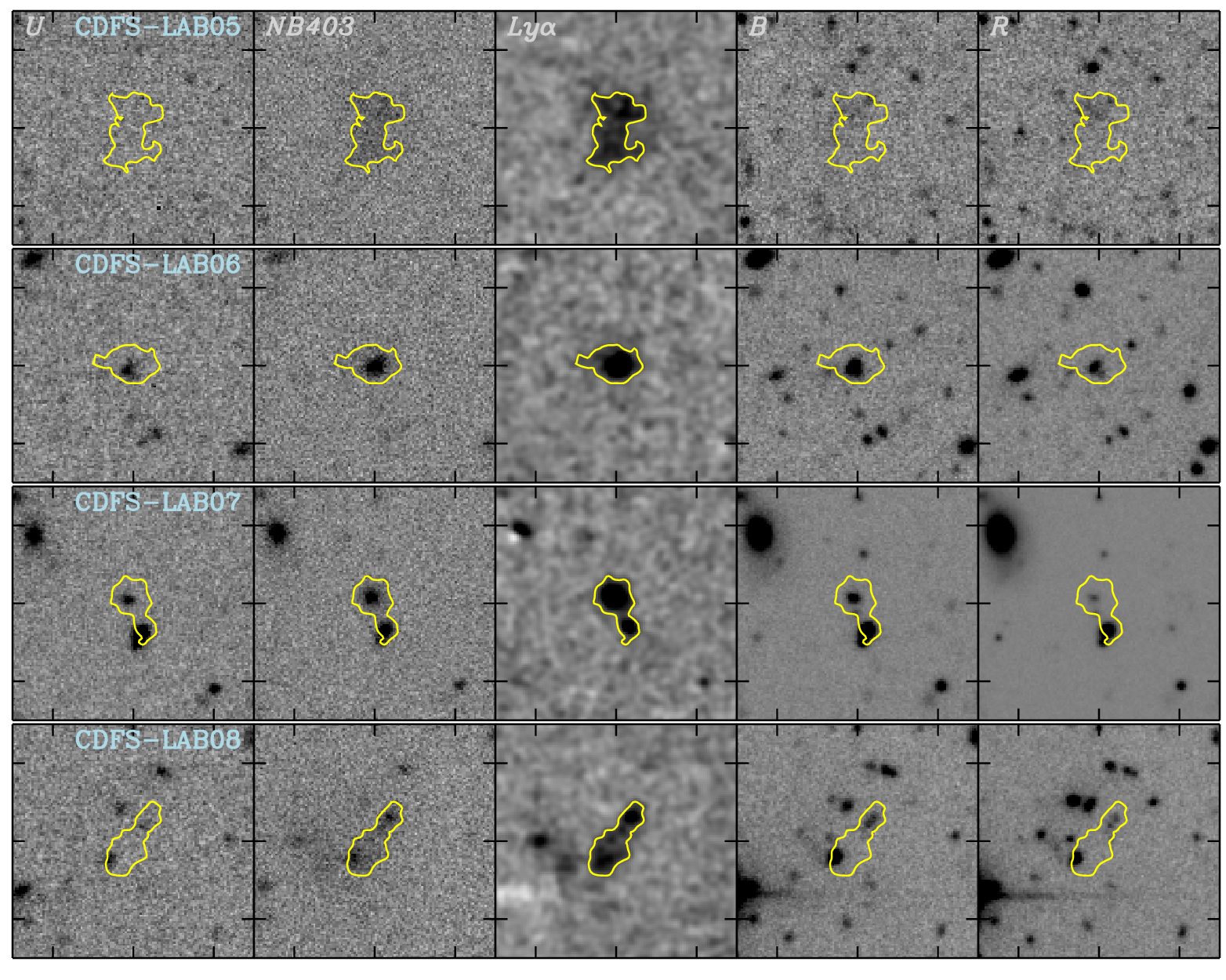}
\caption{Continued.}
\end{figure}


\begin{figure}[!t]
\addtocounter{figure}{-1}
\epsscale{1.2}
\plotone{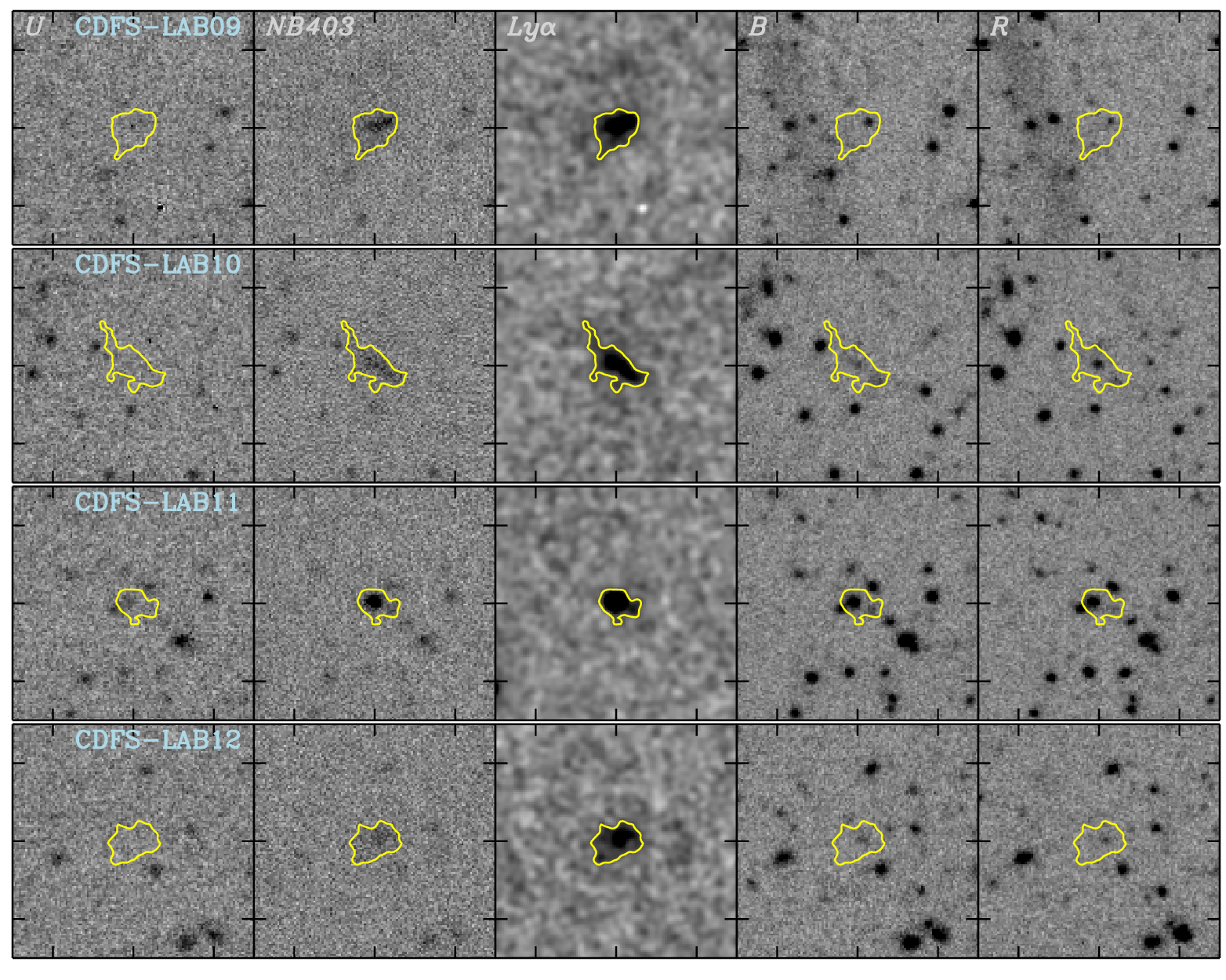}
\caption{Continued.}
\end{figure}


\begin{figure}[!t]
\addtocounter{figure}{-1}
\epsscale{1.2}
\plotone{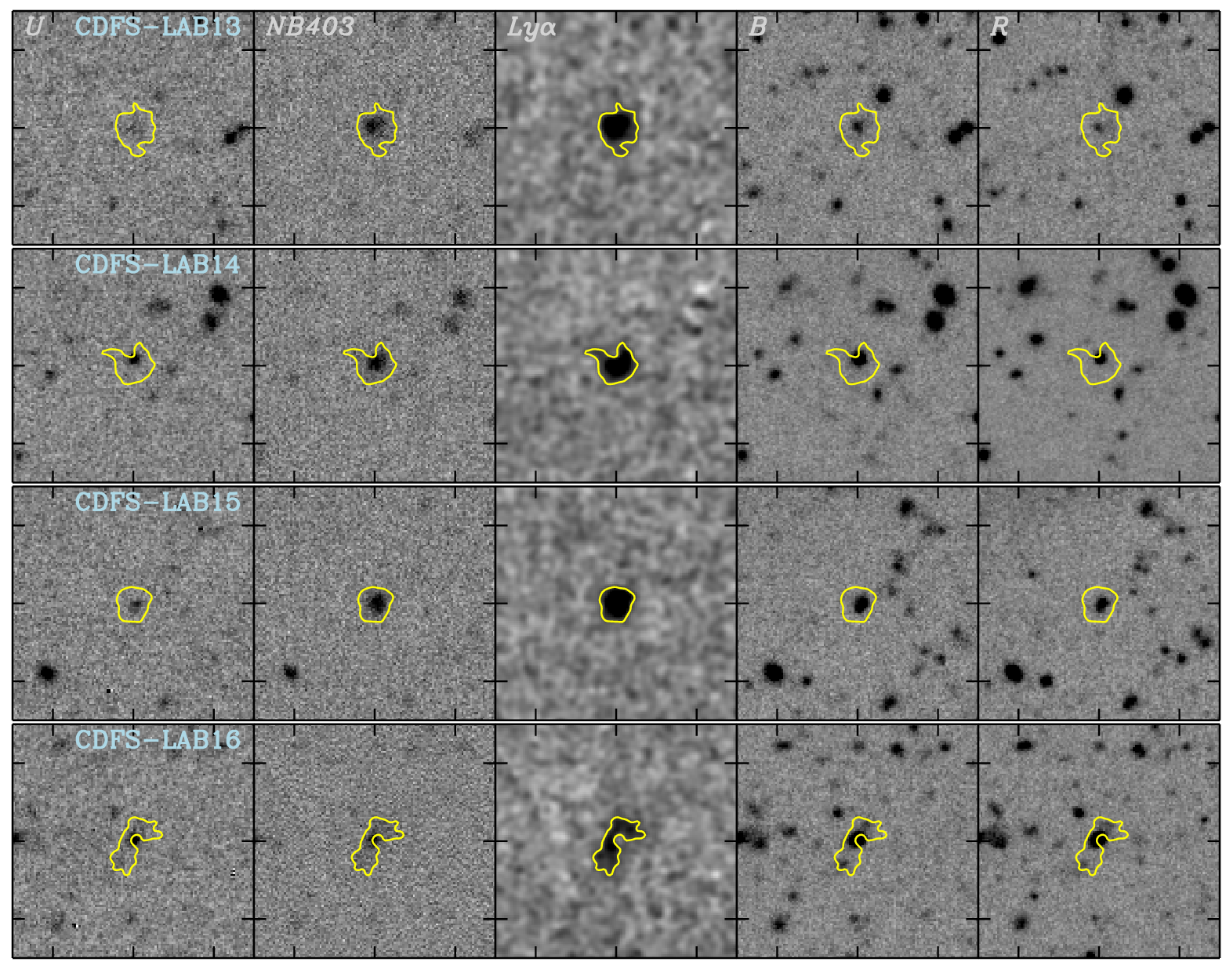}
\caption{Continued.}
\end{figure}


The total area of our survey is $\sim$ 1.2 deg$^2$.  Our \lya\ blob
(LAB) selection criteria, EW$_{\rm obs}$ $>$ 100\,\AA\ and \Aiso $>$
10\,\sq\arcsec\ above the surface brightness threshold of 5.5 $\times$
$10^{-18}$ \unitcgssb, yield a total of 16, 5, 3, and 1 blob candidates
in the CDFS, CDFN, COSMOS1, and COSMOS2 fields, respectively.   We list
their basic properties in Table \ref{tab:properties}:  coordinates,
\lya\ luminosity, and isophotal area.

\begin{figure}[!t]
\epsscale{1.2}
\plotone{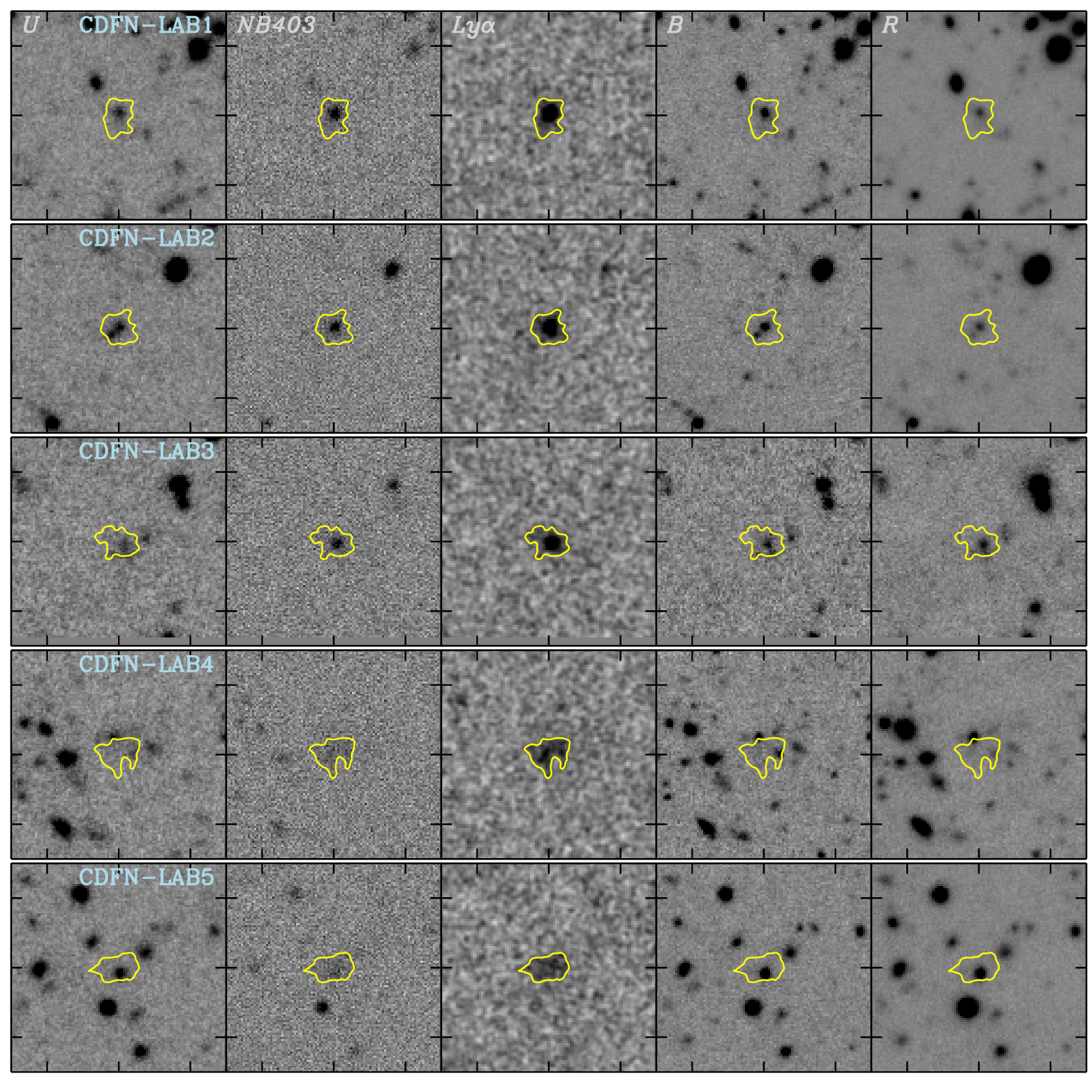}
\caption{
Same as Figure \ref{fig:blob_image_cdfs}, but for the five \lya\ blob
candidates in the CDFN field.
}
\label{fig:blob_image_cdfn}
\end{figure}

\begin{figure}[!t]
\epsscale{1.2}
\plotone{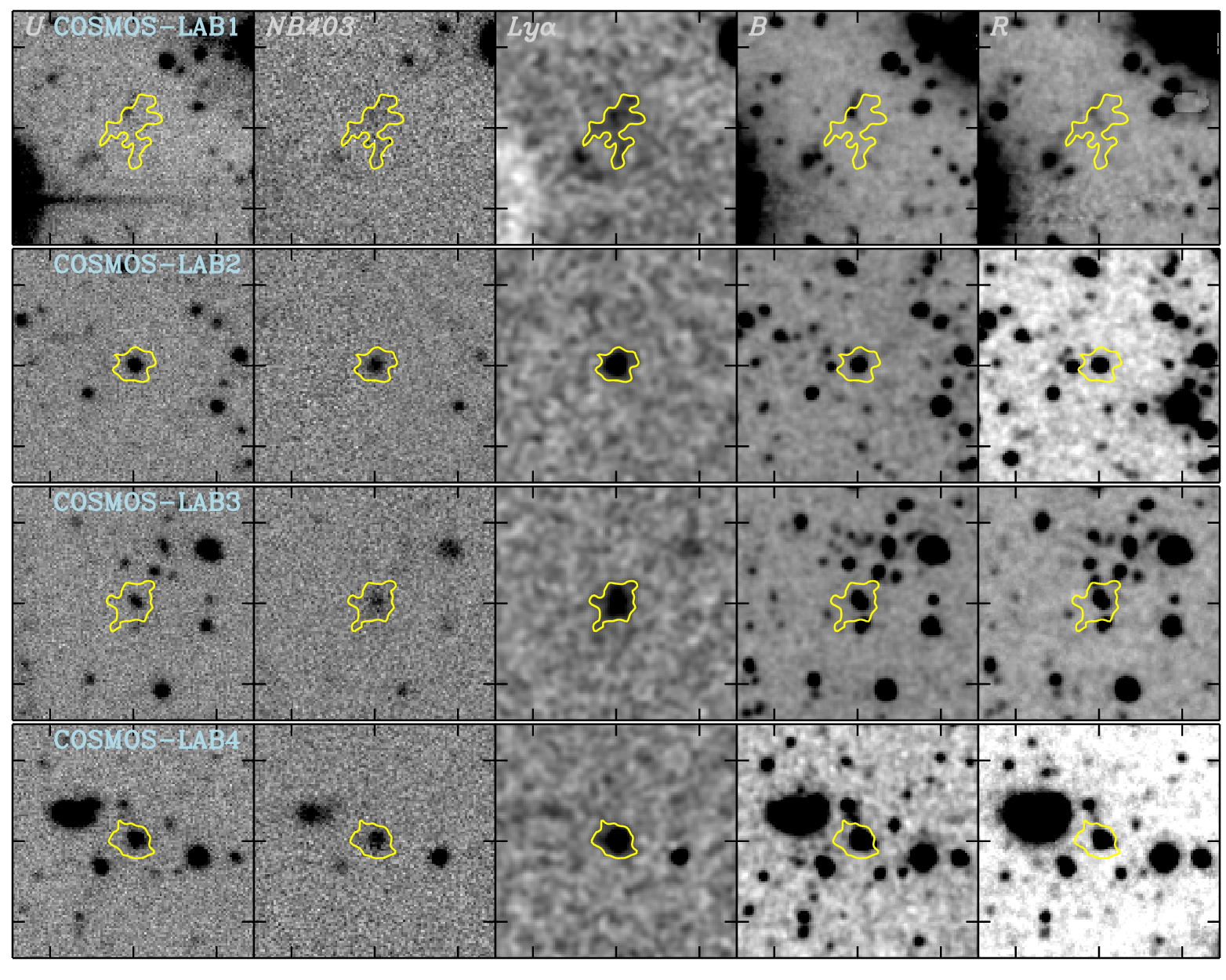}
\caption{
Same as Figure \ref{fig:blob_image_cdfs}, but for the three \lya\ blob
candidates in the COSMOS1 subfield (top rows) and the one candidate in
the COSMOS2 subfield (bottom row).
}
\label{fig:blob_image_cosmos}
\end{figure}

Figures \ref{fig:blob_image_cdfs}--\ref{fig:blob_image_cosmos} show
postage-stamp images of all 25 blob candidates in the \nb403 band,
continuum-subtracted \lya\ line emission, and three broadbands ({\sl UBR})
overlayed with the \lya\ contour corresponding to a surface brightness
of 4 $\times$ 10$^{-18}$ \unitcgssb.

Although we select blob candidates quantitatively, follow-up
visual inspection indicates that the spatial extents of four blobs
are not clearly larger than the local PSF: CDFN-LAB02, CDFN-LAB03,
COSMOS-LAB01, and COSMOS-LAB02. Therefore, we flag them as ``marginal''.
Excluding them does not affect our conclusion regarding the large
field-to-field variation in the blob number density and, in fact, makes
that case stronger.  The largest blobs (more than 16\,\sq\arcsec; the
dashed line in Fig.~\ref{fig:size_luminosity}) are the most robustly
extended and are comparable to those identified in other surveys
\cite[e.g.,][]{Matsuda04}.  Therefore, we treat them separately as the
``bright/large'' subset of the entire blob sample in the clustering
analysis in \S\ref{sec:cosmic_variance}.

The blob candidates have a wide range of sizes and line
luminosities: \Aiso = 10\,\sq\arcsec -- 60\,\sq\arcsec\ and \llya\ =
0.7--8\,$\times$\,$10^{43}$ \unitcgslum. They show diverse morphologies
ranging from compact (the south-west clump of CDFS-LAB01) to diffuse
(CDFS-LAB05) to highly elongated (CDFS-LAB08 and 16).  We do not find
any bubble-like structures that might  be associated with superwinds
like those in a few of the blobs identified by \cite{Matsuda04}.

We are following up the entire blob sample spectroscopically. To date,
we have observed five blobs (CDFS-LAB01, 02, 04, 10, 14) of the 16
in the CDFS.  We confirm all five spectroscopically with their \lya\
and/or H$\alpha$ lines. The details of our spectroscopic campaigns are
presented in a forthcoming paper (Yang et al., in prep.).

The properties of continuum objects associated with blobs provide valuable
clues to the source of blob emission (Yang et al., in prep.). For example,
the blob discovered in the CDFS at $z=3.1$ by \citet{Nilsson06} is not
associated with any continuum source within $\sim$3\arcsec\ \cite[but
see also][]{Geach09}.  Thus, it is possible that this blob is powered by
cooling radiation or cold mode accretion.  Although most blob candidates
in our sample have clear continuum source counterparts in the rest-frame
UV images, two blobs (CDFS-LAB04 and CDFS-LAB05) do not.  We are, however,
able to identify continuum sources within these blobs in deep, rest-frame
{\sl UV} {\sl HST} images \cite[GEMS and GOODS-S;][]{Rix04,Giavalisco04}.
Preliminary inspection of the multiwavelength images also reveals bright
IR or X-ray sources in these blobs.
Interestingly, we often find {\it multiple} continuum sources in a blob,
notably in LAB02 and LAB03.  Our initial examination of the {\sl HST}
images of the 16 \lya\ blobs in the CDFS often resolves these sources as
galaxies, suggesting that star formation and/or nuclear activity might
play a role in producing the \lya\ emission \cite[e.g.,][]{Colbert06}.
This apparent clustering of sources within some blobs may indicate
that blobs are the progenitors of groups or clusters of galaxies today.
We test this possibility quantitatively in \S\ref{sec:cosmic_variance}.



\subsection{Continuous LAE-to-LAB Sequence}
\label{sec:sequence}

There is no evidence of a discontinuity between the properties of our
blobs and unresolved, compact \lya\ emitting (LAE) galaxies (i.e.,
the sources below the solid lines in Fig.~\ref{fig:size_luminosity}).
The smoothness of the LAE-to-LAB transition, which has also been
observed by  \citet{Matsuda04}, suggests that these two types of \lya\
sources may not be distinct and that whatever mechanism or mechanisms
power blobs work over a wide range of luminosity and spatial extent.
Understanding the origin of the extended \lya\ emission requires us
to probe the host galaxy properties --- including the stellar mass,
star formation rate, and size of the star-forming region --- along
the emitter-to-blob sequence.  
If extended star formation or AGN are responsible for powering the
emission, we would expect that the properties of blobs ($A_{\rm iso}$,
$L_{\rm Ly\alpha}$) are correlated with either star formation rate or
X-ray luminosity of the host galaxies.  For example, \citet{Geach05}
argue that a correlation between the \lya\ and bolometric (actually FIR)
luminosity (although weak) suggests that the interaction of an ambient
halo of gas with a galactic-scale superwind is responsible for the
majority of LABs. We discuss the multi-wavelength properties of galaxies
within or near the blobs, considering them as possible energy sources,
in a separate paper (Yang et al., in prep.).

\subsection{Significant Differences in Blob Counts per Field}
\label{sec:cosmic_variance}

Surprisingly, most blobs (16/25) and all eight of the brightest, largest 
blobs (\Aiso $>$ 16\,\sq\arcsec, \llya\,$\gtrsim$ 1.5$\times$10$^{43}$
\unitcgslum) lie in only one of the survey fields: CDFS.  Because the
depth and seeing of the CDFS images are also superior to the other fields,
we first need to verify that this field-to-field variation of the blob
number density does not arise from selection effects.  For example,
although the separation between the fainter/smaller blob candidates and
the simulated point-sources is distinct in the CDFS, the separations in
other three fields are less clear.



To confirm that the field-to-field variation of the blob population is
not due to the different seeing and survey depth, we simulate how
the blob candidates in the CDFS field would look if observed in the
other three fields.  Using the continuum-subtracted postage stamp
(27\arcsec$\times$27\arcsec) images of the CDFS blob candidates, we first
degrade their seeing to that of the CDFN, COSMOS1, and COSMOS2 fields
by convolving kernels derived from PSF images. We rebin the images (if
pixel scale is different), add Poisson noise, and place them into empty
sky regions in the continuum-subtracted narrowband images of the other
fields.  We then measure the blob sizes and luminosities in the same way
as described in \S \ref{sec:sample_selection}.  We repeat this experiment
$\sim$ 1000 times to derive the range of recovered luminosities and sizes.

We show the isophotal areas and \lya\ luminosities (\Aiso -- $L_{\rm
Ly\alpha}$) from this recovery test as green squares in the CDFN,
COSMOS1, and COSMOS2 panels in Figure \ref{fig:size_luminosity}.  If we
estimate the number of possible detections as $N = \sum_{f_{\rm recv}
> 50\%} f_{\rm recv}$, the number of CDFS-like blobs that should have
been detected as bright/large blobs with \Aiso $>$ 16\,\sq\arcsec\
and \llya\,$\gtrsim$ 1.5$\times$10$^{43}$ \unitcgslum\ is 6.9, 6.0, and
6.0, respectively, for each of the three other fields.  However, we do
not find such blobs in other fields, so we conclude that the observed
field-to-field variation is real and that the minimum variation in the
number counts of the largest blobs (\Aiso $>$ 16\,\sq\arcsec) is at least
$N_{\rm blob}$ = ($N_{\rm CDFS}$, $N_{\rm CDFN}$, $N_{\rm COSMOS1}$,
$N_{\rm COSMOS2}$) = (6, 0, 0, 0).  Throughout the rest of this paper,
we refer to this as the ``bright/large blob sample'' and use it as the
default for analyzing blob statistics.  When we lower the criteria for
the blobs to \Aiso $>$ 10\,\sq\arcsec, we expect 14.0, 13.9, and 13.7
blobs in the other three fields, respectively, whereas we find 5, 3, and
1 blob candidates. In this case, the observed contrast between the four
survey fields is $N_{\rm blob}$ = (14, 5, 3, 1).  We subsequently refer
to this as the ``entire blob sample''.  We consider the bright/large
sample and entire sample field-to-field variations in $N_{\rm blob}$,
which are corrected for the different survey conditions, in deriving
the average number density and its variance in the following paragraphs.
The blob statistics are summarized in Table \ref{tab:stat}.


\begin{figure*}[!t]
\plottwo{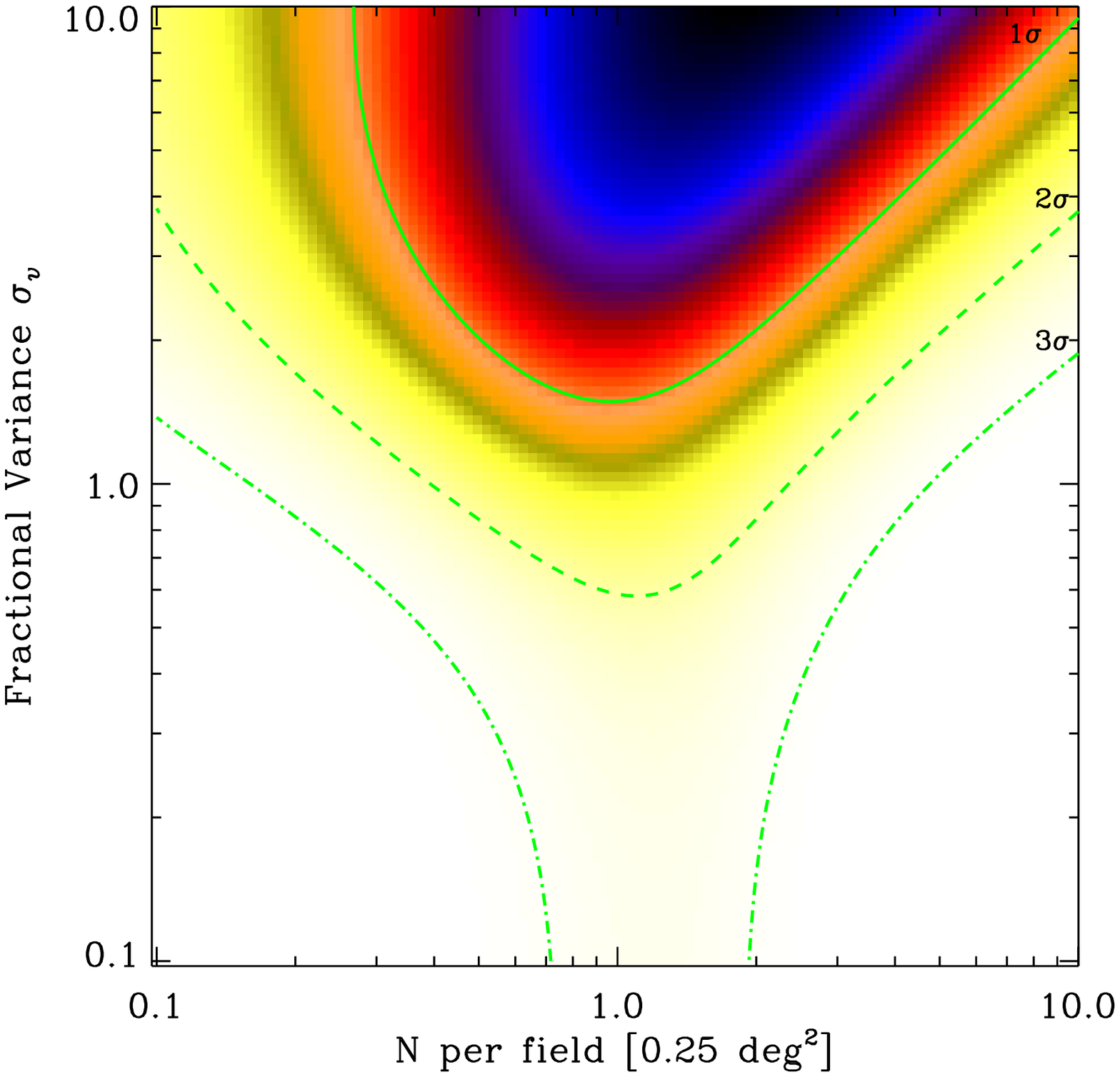}
        {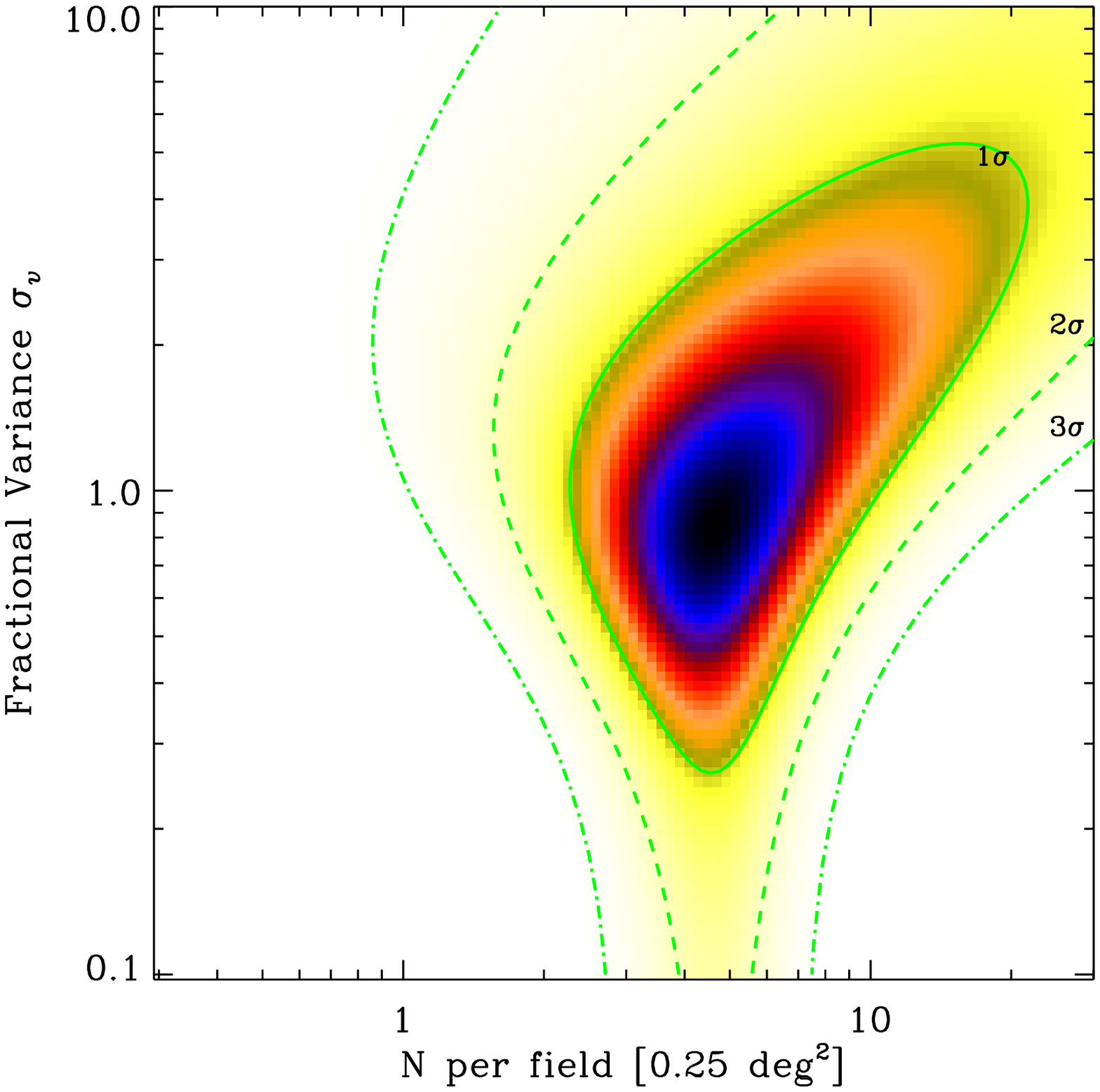}
\caption{
Posterior probability distribution of the fractional variance ($\sigma_v$)
and the average surface density per field ($\bar{N}$) for
the bright, large blobs ({\it left}) and the entire blob sample ({\it
right}).  We adopt logarithmic priors for both $\sigma_v$ and $\bar{N}$.
The 1$\sigma$, 2$\sigma$ and 3$\sigma$ confidence regions are determined
such that the corresponding contours enclose 68.3\%, 95.5\%, and 99.7\% of
the total posterior probability.  The bright, large blobs have fractional
variances of at least 1.5 (150\%) at $1\sigma$ and 0.6 (60\%) at $2\sigma$
regardless of the estimated number density, confirming that blobs are
highly clustered.  For the entire blob survey, the most probable estimate
is ($\bar{N}$, $\sigma_v$) $\simeq$ (4.8, 0.9).
}
\label{fig:variance}
\end{figure*}


Could this observed field-to-field variation arise solely from the
statistical uncertainty? Before we proceed to a detailed analysis, we
have to rule out the possibility that this result arises solely from
Poisson (shot) noise.  For the hypothesis that the surface density
of the bright/large \lya\ blob sample is 2--20 blobs deg$^{-2}$ or
$\sim$ 0.5 -- 5 blobs per survey field, we calculate the probability
of observing non-detections in three survey fields and detections more
than six blobs in any of the fields. We are able to rule out a uniform
distribution with at least 99.97\% confidence ($\sim$3.6$\sigma$). For
the entire blob sample, a uniform distribution is excluded at the
$3.7\sigma$ level.  In \S\ref{sec:halo_mass}, we compare the observed
field-to-field variations with those derived from Poisson statistics and
cosmological N-body simulations, respectively. We show that the latter
better reproduces the observations.

\subsubsection{Quantifying Blob Field-to-Field Variance}
\label{sec:cosmic_variance_est}

In this section, we estimate the field-to-field variation in the number
density of \lya\ blobs from the observed number statistics, $N_{\rm
blob}$. According to $\Lambda$CDM cosmology, the number density and
variance of a galaxy population are not entirely independent properties,
but a function of halo mass.  Here we treat them as independent parameters
and aim to measure them as observables by adopting a simple analytic
approximation of the underlying fluctuations in blob number density
arising from large-scale structure. This method has the advantage that
the blob number density and variance can be derived over a wider range
of parameter space than sampled by simulations.
We then compare these properties with the predictions from the simulations
to obtain constraints on the halo mass (\S\ref{sec:halo_mass}). As
will be shown in the following section (\S\ref{sec:number_density}),
one must consider the field-to-field variations to correctly estimate
the uncertainties in number density and to thus compare blob statistics
across different surveys.

The simplest way of quantifying the field-to-field variation,
$\sigma^2_v$, is to adopt the relation \cite[][\S36]{Peebles80}:
\begin{equation}
\label{eq:variance}
\sigma^2_v = \frac{\avg{N^2} - \avg{N}^2}{\avg{N}^2} - \frac{1}{\avg{N}},
\end{equation}
where $N$ is the number of blobs per 0.25 deg$^{2}$, the typical area
of each of our four fields, and $\sigma^2_v$ is the fractional variance
corrected for Poisson noise.  The number density $n$ can be derived
directly from $N$ by dividing it by the survey volume: $n$ $\equiv$
8.6 $\times$ 10$^{-6}$ $N$ Mpc$^{-3}$.  
Hereafter, we use surface density to mean surface number density.
Thus $\sigma^2_v$ represents
the fractional uncertainty in the observational estimate due to finite
survey volume.
For the bright/large sample, we obtain $\avg{N} = 1.5$ and $\sigma_v$
$\simeq$ 1.5 (150\%).  For the entire blob sample, $\avg{N} = 5.75$
and $\sigma_v$ $\simeq$ 0.76, or 76\%.  Because of our small number
statistics, we choose to adopt a more sophisticated method to better
understand the possible range of blob number density and its variance.

To quantify the field-to-field variance of the blob population, we
calculate the posterior probability for $\sigma^2_v$ and an average
surface density per 0.25 deg$^{2}$ ($\bar{N}$) given our
observation ($D$) of (6, 0, 0, 0) blobs in the bright/large sample:
\begin{equation}
p(\sigma_v, \bar{N}|D) 
\propto 
prob(D| \sigma_v, \bar{N})\ p(\sigma_v,\bar{N}). 
\end{equation}
%
First, we assume that the field-to-field variance follows the log-normal
distribution: $({N}/{\bar{N}})$ $\sim$ {\rm Log--N}$(0, \sigma^2_{\rm
{LN}})$; in other words, $\log ({N}/{\bar{N}})$ follows a normal
distribution $N(0, \sigma^2_{\rm {LN}}$).  Here, $\sigma^2_{\rm LN}$
is the variance of the log-normal distribution and is related to the
actual variance by $\sigma_v^2$ = exp(${\sigma^2_{\rm LN}}$) $-$ 1.
Unlike a Gaussian distribution, the log-normal distribution does not
allow negative values for $N$ and naturally introduces a skewness into
the distribution.  When $\sigma_v \ll 1$, the log-normal distribution is
similar to Gaussian, but the distribution becomes skewed toward zero as
$\sigma_v$ increases, effectively mimicking the dark matter fluctuations
at the high mass end \citep{Coles&Jones91,Bernardeau&Kofman95}.  We choose
the log-normal distribution for the simplicity here, but any reasonable
functional form capable of representing this skewness can be used.

Second, for a given set of ($\sigma_v,\bar{N}$), we calculate the
probability, $prob(D| \sigma_v, \bar{N})$, of finding six blobs in one
field and none in three other fields assuming that the observed number
of blobs follows Poisson statistics with a mean of $\bar{N}$.
We adopt logarithmic priors for both $\sigma_v$ and $\bar{N}$, which
indicates $p(\sigma_v)$ $\propto$ 1/$\sigma_v$ or $p(\bar{N})$ $\propto$
1/$\bar{N}$, implying that the {\it scale} of $\bar{N}$ and $\sigma_v$
is unknown, i.e., that the priors are uniform in logarithmic bins.
We consider a range of $0.1<\bar{N}<30$ and $0.1<\sigma_v<10$, i.e.,
10\% to 1000\% field-to-field variance.  We also test other priors
including (1) a linear prior for both $\sigma_v$ and $\bar{N}$ and
(2) a logarithmic prior for $\sigma_v$ and linear prior for $\bar{N}$,
but the choice of prior does not affect our conclusions.

%

Figure \ref{fig:variance} shows the posterior probability distribution
of the average surface density and variance for the bright/large blob
sample ({\it left}) and the entire sample ({\it right}).
For the bright/large blobs, the posterior favors high variance ($\sigma_v
\gg 1$) as expected.  The confidence regions are not closed, allowing us
to put only a lower bound on $\sigma_v$.  The lower limits are $\sigma_v$
$>$ 1.45 and $\sigma_v$ $>$ 0.57 for the 1$\sigma$ and 2$\sigma$
confidence levels, respectively, for the joint distribution (i.e., we
attempt to constrain both $\bar{N}$ and $\sigma_v$ at the same time).
The 1$\sigma$, 2$\sigma$ and 3$\sigma$ confidence regions are determined
such that the corresponding contours enclose 68.3\%, 95.5\%, and 99.7\%
of total posterior probability.  The 3$\sigma$ lower limits are not
constrained because the priors become too high as $\sigma_v \rightarrow
0$, thus the posterior probability is determined more by the input priors
rather than by the data themselves.

We also calculate the confidence interval for each parameter ($\sigma_v$
or $\bar{N}$) by marginalizing out the other parameter in the posterior
probability distribution: e.g., $p(\bar{N}|D) = \int p(\sigma_v,
\bar{N}|D) d\sigma_v$. This interval then represents the estimate of one
parameter independent of the other.  In \S\ref{sec:number_density}, we
compare the marginalized surface density estimates, {\it regardless of the
variance}, with those obtained from past observational studies.  In Table
\ref{tab:cosmic_variance}, we show the blob surface density ($\bar{N}$)
and variance estimates obtained from the above.  The field-to-field
variation of the bright/large blobs is stronger than 150\% and 60\%
at the 1$\sigma$ and 2$\sigma$ confidence levels, respectively.
For the entire blob sample, the most probable parameters are ($\bar{N}$,
$\sigma_v$) $\simeq$ (4.8, 0.9) with large uncertainties of $0.26 <
\sigma_v < 5.2$ at the 1$\sigma$ level.  The large variance implies
a highly skewed distribution and thus reproduces the observed strong
field-to-field variations.

Observationally, $\sigma_v$ $\sim$ 1.5 (150\%) is much larger than for
compact \lya\ emitters (LAEs) at $z$ = 3--5 obtained from narrowband
imaging over volumes comparable to our survey.
For example, we estimate a LAE variance of $\sim$ 20--30\% at
z$\approx$3.1 from the five 0.2 deg$^2$ subfields of \citet{Ouchi08}.
From \citet{Shioya09}, we estimate that the LAE variance at $z\approx4.86$
is likewise $\sim$ 30\% when their contiguous survey area ($\sim$
2 deg$^2$) is divided into 0.25 deg$^2$ subfields (their Table 2).
This $\sim$30\% variance is enough to produce the factor of 2 difference
among the LAE number densities observed in their fields.  Note that the
$\sim$30\% LAE variance is uncorrected for Poisson noise, which would
lower it.
Therefore, the even larger variance of blobs suggests that they lie in
halos more massive than those of LAEs.  We put the first constraints on
the dark matter halo mass of blobs in \S\ref{sec:halo_mass} by comparing
our blob statistics with N-body simulations.

The strong variation in blob counts from one 0.25\,deg$^2$ field to
another is consistent with the discovery of a close pair of blobs by
\citet{Yang09}. Those blobs were among only four detected in our shallower,
but larger (4.8\,deg$^2$) survey of the NOAO Bo\"otes field.  They are
separated by only $\sim$70\arcsec.  If that survey had been conducted by
mosaicing 30\arcmin$\times$30\arcmin\ fields like those sampled by our
imager here, most fields would not contain the pair.  On a related note,
the clustering of 35 blobs in the SSA22 overdensity \citep{Matsuda04}
was identified during a follow-up of two giant blobs originally found
by \citet{Steidel00}.

\subsubsection{Blob Number Density and Luminosity Function}
\label{sec:number_density}

The strong field-to-field variation of the \lya\ blobs presents challenges
for the measurement of their number density and luminosity function
(LF).  In this section, we compare our LFs derived from each survey
field with each other and with previous studies.  We demonstrate that a
large volume survey and/or multiple pointings are critical to constrain
the blob LF.  For the rest of paper, we adopt the marginalized number
densities from the previous section, which give us the number densities
of $n$ = $1.0^{+1.8}_{-0.6}$$\times$$10^{-5}$ Mpc$^{-3}$ (from $N$
= 1.2 per 0.25 deg$^2$) for the bright/large blob sample and $n$ =
$4.1^{+4.8}_{-1.6}$$\times$$10^{-5}$ Mpc$^{-3}$ (from $N$ = 4.8 per 0.25
deg$^2$) for the entire blob sample.


\begin{figure}[!t]
\epsscale{1.2}
\plotone{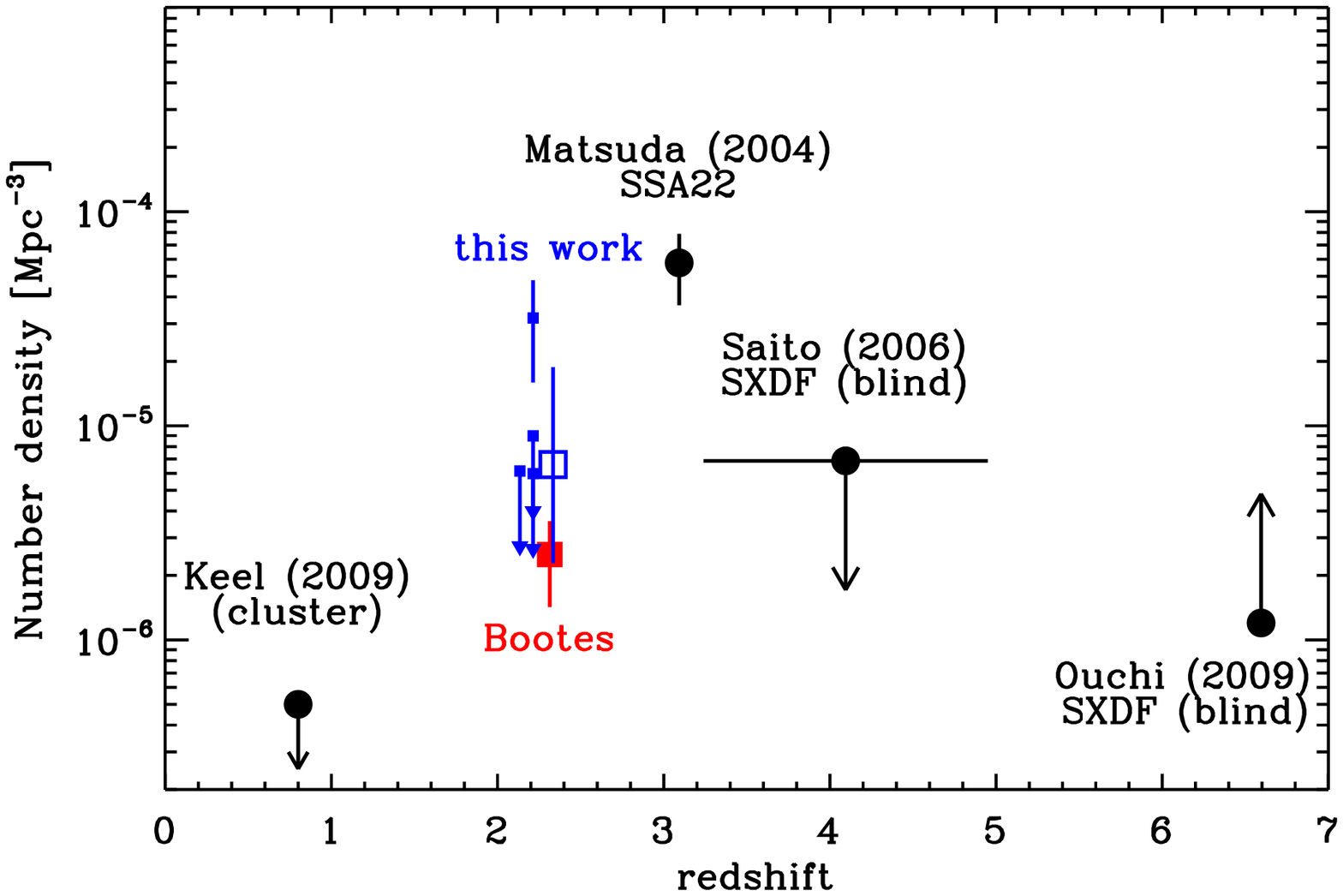}
\caption{
Number density of \lya\ blobs at different redshifts.  The filled and
open squares are from our two narrow-band imaging surveys at $z=2.3$:
the ultra-wide Bok+90Prime survey in Bo\"otes \citep{Yang09}
and the narrower, but deeper, NOAO-4m survey (this work),
respectively. The filled dots are the number density estimates from
\citet{Matsuda04,Saito06,Ouchi09,Keel09}.  All the number densities
plotted here are scaled to match the shallowest survey (Bo\"otes
field) with $\sim$1.8\arcsec\ seeing and an rms sky background of
5$\times$10$^{-18}$ \unitcgssb.  The upper or lower limits on several
of the points from the literature are explained in \citet{Yang09}.
The four small squares represent the individual number density estimates
for each of our four survey fields, demonstrating that a large volume
survey is required to overcome the strong field-to-field variation of
the blobs.  Note that these data points are shown slightly shifted for
clarity, but they are all at $z=2.3$.
}
\label{fig:number_density}
\end{figure}


First, we compare the number density of bright/large blobs derived
from \S\ref{sec:cosmic_variance_est} with previous measurements (Figure
\ref{fig:number_density}).  To make a fair comparison with other work, we
need to cut the bright/large sample to satisfy the selection criteria used
by \citet{Yang09} when they previously examined the blob number densities
among different samples spanning $z$ = 0.8 to 6.6.  Those criteria were:
\llya\ $>$ 1.5$\times$10$^{43}$ \unitcgslum\ and \Aiso $>$ 25\,\sq\arcsec\
above a surface brightness threshold of 5$\times$10$^{-18}$ \unitcgssb\
under $\sim$1.8\arcsec\ seeing. We refer readers to \citet{Yang09}
for the details about the estimates at each redshift.

Only four blobs (CDFS-LAB01, 02, 03, and 04) from the sample of six
bright/large blobs satisfy the \citet{Yang09} criteria, so we scale the
bright/large counts down by a factor of 4/6, obtaining a surface density
of $0.77^{+1.4}_{-0.5}$ per 0.25\,deg$^2$ fields or a number density of
$n$ = $0.66^{+1.2}_{-0.4}$ $\times$10$^{-5}$ Mpc$^{-3}$ ({\it large,
open square} in Fig.\ \ref{fig:number_density}).  The error bar takes
into account the field-to-field variation as well as the Poisson noise.
This average number density at $z=2.3$ derived from the four NOAO-4m
survey fields is thus consistent with the \citet{Yang09} measurement at
the same redshift ($n$ = $0.25{\pm0.10}$ $\times$10$^{-5}$ Mpc$^{-3}$;
{\it large, filled square}) obtained from our $\sim$ 4$\times$ larger
volume survey of Bo\"otes.

To illustrate the uncertainties arising from the field-to-field
variance, we show the bright/large number density for each of the four
fields individually.  The small squares represent $n$ = 3.2\,($\pm$1.6)
$\times$10$^{-5}$ Mpc$^{-3}$ for the CDFS, and three upper limits, $n$
$<$ 0.90, 0.62, 0.60 $\times$10$^{-5}$ Mpc$^{-3}$ for CDFN, COSMOS1,
and COSMOS2, respectively.  The number density in the CDFS field is
consistent with that found by \citet{Matsuda04} in the SSA22 field
(5.8$\pm$2.1\,$\times$10$^{-5}$) within the uncertainties, suggesting 
that blobs at $z = 2.3$ in the CDFS also lie in an overdense region
\cite[see also][]{Palunas04,Prescott08} and occupy high mass halos.
On the other hand, the blob number densities for the other three fields
are lower than for the CDFS and the SSA22 field, demonstrating that
characterizing the evolution in $n$ with $z$ requires surveys large enough
to overcome the field-to-field variance.  Because such measurements are
not yet available at other redshifts, it is not possible to constrain
$n(z)$ at this time.

Note that we are not able to apply the \citet{Yang09} selection criteria
to the higher-$z$ \citet{Saito06} and \citet{Ouchi09} blob samples
because those authors use selection methods different than ours and
\citet{Matsuda04}.  Therefore, we plot these higher-$z$ estimates as
upper and lower limits, respectively.  Unlike the comparison between
the $z=2.3$ and $z=3.1$ samples, comparison of the \citet{Saito06} and
\citet{Ouchi09} points with our results is perilous. We plot all these
points together only to summarize the state of blob surveys.

In addition to the blob statistics obtained from narrow- or
intermediate-band imaging techniques, we show the number density estimate
($n$ $<$ 0.5$\times$10$^{-6}$ Mpc$^{-3}$) from GALEX slitless spectroscopy
of two galaxy (super)clusters at $z\simeq0.82$ \citep{Keel09}. This
$n$, even though calculated from overdense regions, is well below
that of the overdense CDFS and \citet{Matsuda04} fields, supporting the
\citet{Keel09} claim that \lya\ blobs might be high redshift phenomena.
Additional surveys at $z < 1$ are required to confirm this result, as we
do not know whether the variance in $n$ at $z=2.3$ persists at lower $z$.


\begin{figure}[!t]
\epsscale{1.10}
\plotone{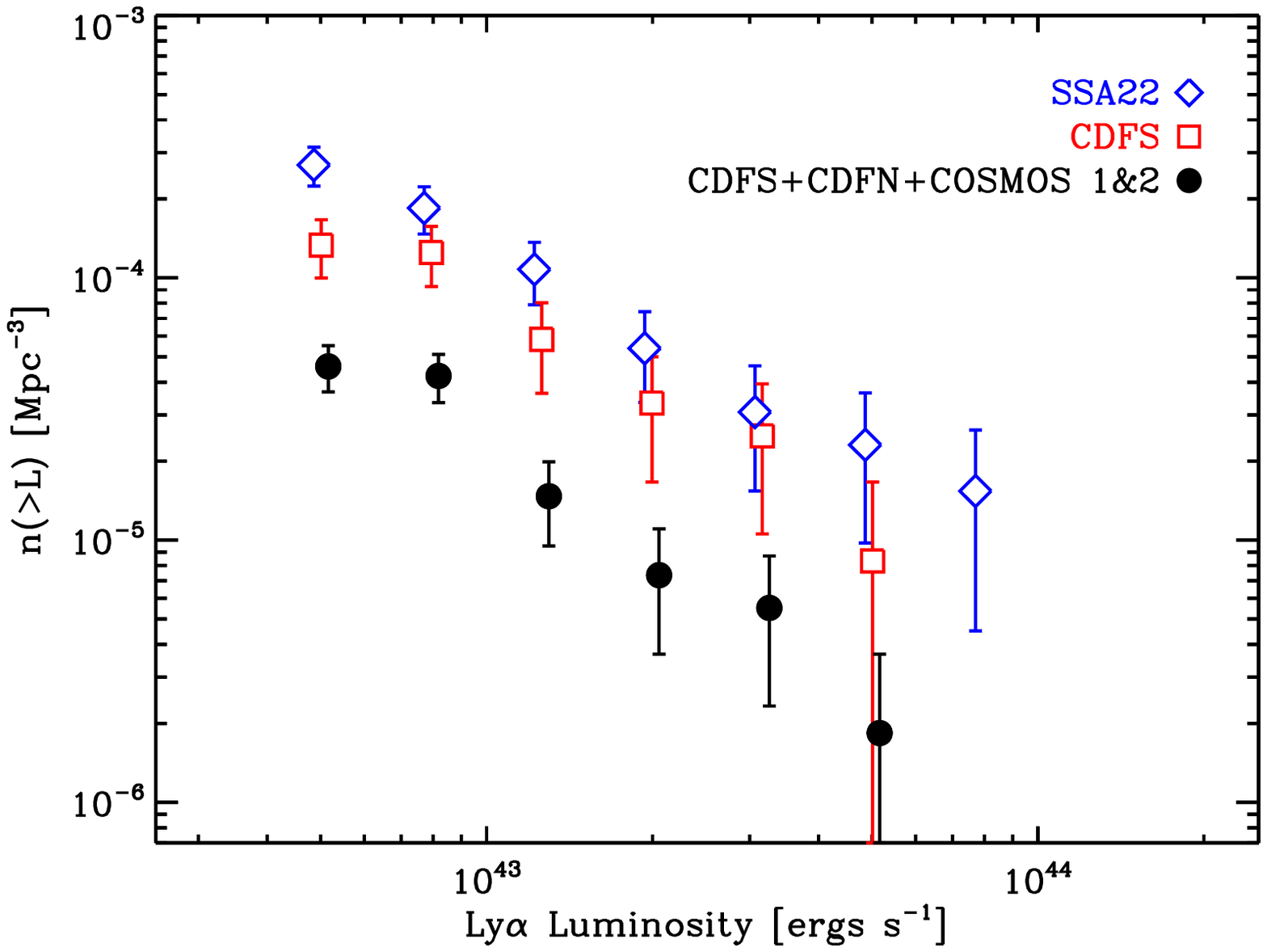}
\caption{
Cumulative \lya\ luminosity function (LF) for \lya\ blobs. The open
squares (red) and diamonds (blue) represent the LFs derived from the
CDFS (this paper) and SSA22 \citep{Matsuda04} fields, respectively,
which each have a survey area of $\sim$0.25\,deg$^2$.  The filled dots
represent the average LF measured from all four of our survey fields
(CDFS, CDFN, COSMOS1, and COSMOS2). Data points are slightly shifted
relatively to each other for clarity.  While the slope and normalization
of the CDFS and SSA22 LFs roughly agree, the normalization of the average
LF is $\sim$ four times lower, suggesting that one should account for
the strong field-to-field variation of \lya\ blobs when calculating LFs.
}
\label{fig:lab_lf}
\end{figure}


Second, we compare our blob luminosity function with that
of \citet{Matsuda04} in Figure \ref{fig:lab_lf}. The similar
depth and selection criteria of the two surveys makes the direct
comparison meaningful (see Appendix \ref{sec:blob_comparison}). For
each of our blobs, we use the {\it total} \lya\ luminosity (Table
\ref{tab:properties}), which our recovery test demonstrates is reliable.
The isophotal flux could be biased due to the different signal-to-noise
and filtering kernels used in the detection procedure.  We construct
an average LF as follows.  In each \lya\ luminosity bin, we divide the
number of blobs found in all four of our survey fields by the total
survey volume.  We also plot the LF obtained from only the 16 blobs in
the CDFS.  We do not apply a completeness correction using the recovery
fraction $f_{\rm recv}$ in Table \ref{tab:properties}, because we also
do not know the completeness of \citet{Matsuda04} sample.  In Table
\ref{tab:luminosity_function}, we list our blind survey LFs as well as
that of the SSA22 overdense region \citep{Matsuda04}.

The slope and normalization of the CDFS (squares) and \citet{Matsuda04}
(diamonds) LFs agree roughly.  The apparent discrepancy in the faintest
bin ($\log$ \llya = 42.8) is likely due to incompleteness in our survey.
We also do not find any blobs as bright as those in their brightest
bin ($\log$ \llya = 44.0).  Not surprisingly, the normalization of our
``average'' luminosity function (filled circles) is $\sim4\times$ lower
than those of the CDFS and \citet{Matsuda04} LFs.  Once again, we see
that comparisons among surveys are difficult without knowledge of the
blob clustering strength and that large volume surveys are required
to overcome the strong field-to-field variance of the blob counts.
The same conclusion is reached from the on-going Subaru survey of blobs
\cite[see Fig.\ 13 of][Y.\ Matsuda in preparation]{Goerdt09}.

The clustering of blobs provides a means to discriminate among models
for the origin of the extended \lya\ emission, including photoionization
by AGN, outflows due to intense star formation, and cooling radiation.
This analysis is beyond the scope of this paper, but we note that
any blob-producing mechanism must reproduce both the observed LF and
its field-to-field variation.  For example, it would be interesting to
ascertain whether AGN or sub-millimeter galaxies (SMGs) have clustering
strengths similar to blobs.  Along these lines, \citet{Dijkstra&Loeb09}
claim that \lya\ blobs are cooling radiation arising from cold streams
falling onto the embedded galaxies and that a strong variance naturally
arises from the underlying variation of dark matter halos \cite[see
also][]{Goerdt09}.

\subsubsection{Blob Halo Masses}
\label{sec:halo_mass}

Because our blob survey is blind and sufficiently large to determine the
blob number density and its variance, we have a unique opportunity to
constrain the properties of the dark halos in which the blobs reside, and
thus to understand what these mysterious objects have evolved into today.
For example, based on the blob number density and the discovery of the
blob pair in the NOAO Deep-Wide Bo\"otes field, \citet{Yang09} suggest
that blobs are sites for the formation of the brightest galaxies in rich
galaxy clusters.  However, the small number statistics of that study
precluded constraining the mass of blob halos.
Although our current survey statistics are still not large enough
to directly measure the clustering of blobs via correlation function
analysis, we can use $n$, $\sigma_v$, and a cosmological N-body simulation
of $\Lambda$CDM universe to identify the most likely halo mass occupied
by blobs.

We first select a dark halo (DM) mass in which blobs could reside.
All halos above this minimum mass $M_{\rm min}$ have a fixed probability
of containing a detectable blob, which is labeled the detectability
fraction $f_{\rm D}$\footnotemark.
\footnotetext{Note that $f_{\rm D}$ is a statement about the detectability
of the ensemble of halos and is {\it not} the classically-defined
duty-cycle, which is simply the fraction of time a blob is on.}
We choose $f_{\rm D}$ such that the halo mass function from
the simulation reproduces the observed blob number density from
\S\ref{sec:cosmic_variance_est}.  Once the halo mass and detectability
fraction are fixed, we can predict the clustering of such halos,
i.e., the field-to-field variation, directly from the simulation using
counts-in-cells (c-in-c) methodology.  Then we compare this prediction to the
observed variation in blob counts over the four survey fields to quantify
the likelihood that the selected halo mass and detectability fraction
reproduce the observed blob statistics in Table \ref{tab:stat}.

%

To link the number density and its variance to the DM halo mass,
we employ a simple counts-in-cells analysis using the halo catalog
at $z=2$ derived from the ABACUS N-body code (Metchnik \& Pinto,
in prep.).  This simulation has a cubic volume of 1$h^{-1}$ co-moving
Gpc on a side and 1024$^3$ dark matter particles with $m_{\rm DM}$ =
1.1$\times$10$^{11}$$M_{\sun}$.
We adopt the cosmological parameters: $H_0 = 0.701$, $n_s = 0.96$,
$\Omega_{\rm M}=0.279$, $\Omega_b=0.0456$ and $\sigma_8 = 0.817$.
Dark halos are defined using a friends-of-friends algorithm with linking
length $b=0.16$ in units of the initial particle spacing.  The smallest
halos used in our analysis consist of 48 particles.
Due its large size, this simulation is finely tuned to our problem,
which requires sampling many ``cells,'' $\sim$50 comoving Mpc boxes
that are roughly the same size and geometry as our survey of each of
the four fields.

To constrain the DM halo mass of the blobs, we first consider the observed
bright/large blob number density $N$ = 1.2 per 0.25 deg$^{2}$ field,
or $n$ = 1.0$\times$10$^{-5}$ Mpc$^{-3}$ (1.5 blobs in a 50 Mpc box).
If all halos contain a detectable blob (i.e., the detectability fraction
$f_{\rm D}$ is 100\%), then this number density requires halos with more
than 150 DM particles or 1.7$\times$10$^{13}$\,$M_\sun$.
We derive the counts-in-cells distribution of the simulated blobs
by counting the number of blobs within 10000 randomly placed
50\,Mpc boxes and by assuming a simple halo occupancy distribution
\cite[e.g.,][]{Berlind&Weinberg02} with $N_g ({M\geqslant M_{\rm min}})
= 1 + ({M}/{M_1})^\alpha$, where $M_{\rm min}$ = 150 DM particles
or 1.7$\times$10$^{13}$\,$M_\sun$, ${M_1}$ = 1000 DM particles or
1.1$\times$10$^{14}$\,$M_\sun$, and $\alpha = 1$ as a fiducial value.
This cell size is similar to our survey volume,
48.7$\times$48.7$\times$46.8 Mpc, for the 0.25\,deg$^2$ FOV with the
\nb403 narrowband filter.  While the four survey fields have slightly
different survey dimensions, the choice of a 50\,Mpc box does not affect
our conclusions.

\begin{figure}[!t]
\epsscale{1.1}
\plotone{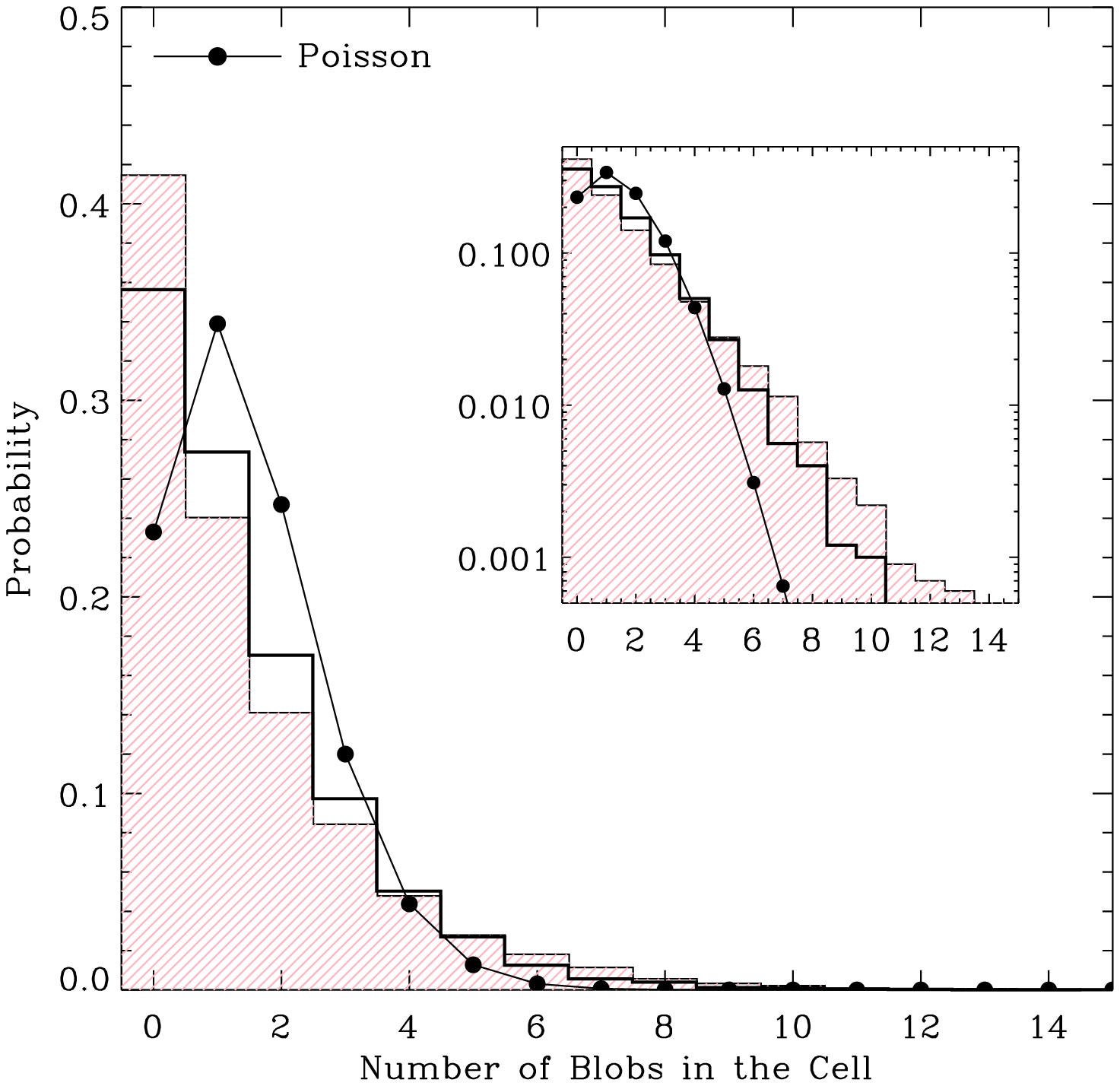}
\caption{
Counts-in-cells distribution of blobs drawn from the ABACUS 1
$h^{-1}$Gpc N-body simulation (Metchnik \& Pinto, in prep.) ({\it
shaded histogram}) that have the same average number density as our
bright, large blob sample ($n$ = 1.0$\times$10$^{-5}$ Mpc$^{-3}$).
We count the number of blobs within 10000 randomly placed 50\,Mpc
boxes assuming that all massive halos of at least $M_{\rm min}$ =
1.7$\times$$10^{13}$\,$M_{\sun}$ have detectable blobs (i.e., the
detectability fraction is 100\%).  The simulated distribution has a
Poisson-noise corrected variance $\sigma_v$ = 1.04, consistent with our
estimates from \S\ref{sec:cosmic_variance_est}. The thick solid histogram
represents the distribution for a detectability fraction of 12.5\%
and a blob halo mass of $M_{\rm min}$ = 0.5$\times$$10^{13}$ $M_{\sun}$.
The observed number density and the strong field-to-field variation of
the bright, large blobs at $z=2.3$ suggest that they reside in halos that
are likely to evolve into the $\sim 10^{14}$\,$M_{\sun}$ halos typical
of present-day clusters of galaxies.  The dots represent the Poisson
noise with a mean of 1.46.  The inset shows the same distribution but
for a logarithmic scale.
}
\label{fig:counts-in-cells}
\end{figure}

Figure \ref{fig:counts-in-cells} shows the simulated bright/large blob
counts-in-cells distribution with an average of 1.46 blobs per cell.
The overlayed line represents a Poisson distribution with the same
average, a reference case in which the DM halos containing blobs are not
clustered.  Thus, comparing the bright/large and Poisson distributions
tests the null hypothesis that the halos and blobs are not correlated.

As expected from the large field-to-field variation
(\S\ref{sec:cosmic_variance_est}), the low and high tails of the simulated
distribution exceed the Poisson counts.  Using the Eq.~(\ref{eq:variance}),
we derive the variance of this distribution, $\sigma_v = 1.04$, which is
consistent with our estimate ($\sim$1.4$\sigma$ lower limit) from the previous
section (\S\ref{sec:cosmic_variance_est}).  This variance corresponds to a
bias of $b$ = $\sigma_{\rm blob}/\sigma_{\rm DM}$ $\sim$ 7 given the rms
fluctuation of mass ($\sigma_{\rm DM} = 0.15$) within a sphere of 31 Mpc
radius with the same volume as our survey box.


Is this simulated distribution consistent with the observed
field-to-field variation in $n$?  For $f_{\rm D}$ = 100\% (shaded
histogram), the probability of finding no blobs in a cell is as high as
41.5\% ($\sim2\times$ greater than the Poisson probability), while the
probability of finding 6 or more blobs is 4.4\% ($\sim11\times$ greater
than Poisson).  The posterior probability of finding at least six blobs
in one survey field with non-detections in three other fields (the case
for the bright/large blobs in our sample) is 1.3\% (4 $\times$0.415$^3$
$\times$ 0.044).  This posterior probability is $\sim$\,65 times larger
than 0.02\% for the Poisson distribution.  Therefore, we reject the null
hypothesis that the dark matter halos and blobs are uncorrelated.

Although the posterior probability is much higher than Poisson, is
it statistically acceptable as a halo mass constraint?  To test if we
can reject the posterior probability of 1.3\%, we consider an extreme
case that maximizes our posterior statistics:  $P$ = $4 p^3 (1-p)$, the
probability of not finding any blob in three survey fields, but finding
any number of blobs in the one remaining field.  Here, $p$ represents
the probability of finding zero blobs in one survey field, and $P$
has a maximum value of 42.2\% when $p$ = 0.75.  Therefore, we should
compare our posterior probability of 1.3\% with this extreme case, not
with 100\%. The posterior probability is only $\sim$34$\times$ smaller
than this {\it maximum} probability.  Therefore, we conclude that our
measurements are consistent with the halo model assuming $M_{\rm halo}$
$\sim$10$^{13}$ $M_{\sun}$ and $f_D$ = 100\%.

We have now established that the observed strong field-to-field variation
in blob counts is not surprising if massive dark matter halos with $M_{\rm
halo}$ $\gtrsim$ $10^{13}$ $M_{\sun}$ always produce detectable \lya\
blobs.  Here we investigate whether it is possible to obtain consistency
with lower values of halo mass by changing the blob detectability fraction.
We derived the {\it maximal} halo mass for the observed blob number
density by assuming that all halos more massive than $M_{\rm halo}$
$\sim$ 1.7$\times$10$^{13}$\,$M_\sun$ have detectable blobs.  If we
lower $f_{\rm D}$ to 50\%, 25\%, and 12.5\%, while increasing the
number density of halos capable of hosting blobs by 2, 4, and 8 times to
keep the abundance of the observed blobs constant, the threshold halo
mass $M_{\rm min}$ decreases to 1.2, 0.8, and 0.5 $\times$ 10$^{13}$
$M_{\sun}$, respectively.  At lower halo mass, we naturally obtain a
weaker field-to-field variation and a less prominent high-end tail in
the counts-in-cells distribution (i.e., at $N > 6$) than for $f_{\rm D}$
= 100\%.  To illustrate this trend, we also show the counts-in-cells
distributions for $f_{\rm D}$ = 12.5\% (thick solid line) in Figure
\ref{fig:counts-in-cells}.  Table \ref{tab:duty_cycle} summarizes the
counts-in-cells statistics, the field-to-field variance ($\sigma_v$),
and the posterior probability of observing 6 or more blobs in only one
of the survey fields for the different $f_{\rm D}$ values.
For comparison, we also list the counts-in-cells statistics and the
posterior probabilities for the Poisson distributions.


Higher detectability fractions (and more massive halos) produce
a field-to-field variance about the observed $n$ that is more
consistent with the observed variance.  Halo models with $\gtrsim$
$10^{13}$\,$M_{\sun}$ halos and $f_{\rm D} \gtrsim 50\%$ work best.  Lower
detectability fractions (e.g., $\sim$ 12\%) require somewhat lower halo
masses ($\sim$ $5 \times 10^{12}$\,$M_{\sun}$) to reproduce $n$, but the
resulting variance in $n$ is lower and further from the observed value.
However, the effects of lowering the detectability fraction and halo
mass are not large enough to put strict lower limits on the halo mass:
for the lowest $f_{\rm D}$ considered, $12.5\%$, we predict $\sigma_v$ =
0.76 (76\%) and a posterior probability of 0.45\%, which is only $\sim$
3$\times$ lower than in the $f_{\rm D} = 100\%$ case.  It is possible,
but less likely, that the blobs occupy a few $\times 10^{12}$ $M_{\sun}$
halos if $f_{\rm D}$ is much lower than $12.5\%$.

A detectability fraction of $f_{\rm D} = 12.5\%$ implies a short blob
lifetime, only $\tau \lesssim 350$ Myrs at $z=2-3$.  In principle,
we could adopt still lower values ($f_{\rm D} \ll 10\%$), down to the
limit where the \lya\ blobs live only a few tens of Myrs, and thus lower
halo masses ($\lesssim$10$^{12}$$M_{\sun}$), in order to find the point
at which we can reject the assumed halo mass.  However, the limited mass
resolution of our N-body code does not allow us to resolve smaller halos.
Therefore, to put tighter constraints on the halo mass requires improving
the blob statistics by extending surveys to larger volumes and/or creating
higher resolution simulations.

For now, we conclude that bright/large \lya\ blobs are most likely to
reside in massive dark halos with $\gtrsim$ $10^{13}$\,$M_{\sun}$ that
have detectable blobs more than $\sim 50\%$ of the time.  Interestingly,
these halo mass estimates agree with the dynamical masses, $M_{\rm dyn}$
= $10^{12}$--$10^{13}$\,$M_{\sun}$, derived from the width of \lya\
lines in similar blobs \citep{Matsuda06}.  However, special caution
is required in using the \lya\ line width as a mass proxy, because the
radiative transfer effects on the line width are poorly understood.

For the entire blob sample (with counts of 14, 5, 3, 1 for the four survey
fields), the required halo mass for $f_{\rm D}$ = 100\% is $M_{\rm min}$
= 0.8$\times$10$^{13}$\,$M_{\sun}$.  
The fractional variance is $\sigma_v = 0.76$, again consistent with
the observed value.  Because the halo mass function is steep at the
high mass end, $M_{\rm min}$ for the entire sample is similar to that
derived for the bright/large blobs alone.  Therefore, the resulting halo
mass is insensitive to the blob selection cuts, and the statistics for
our entire sample are consistent with blobs occupying halos of $\sim$
10$^{13}$\,$M_{\sun}$.

Halos of $\sim$ 10$^{13}$\,$M_{\sun}$ lie in the high mass tail of the
halo mass distribution at $z = 2.3$.  How massive will these halos be
today?  Because our simulations were not run beyond $z=1$, we consider
the N-body simulation of \citet{Maccio08}, which extends to $z=0$.
Their $M_{\rm halo}$ $\gtrsim$ $10^{13}$ $M_{\odot}$ halos grow in
mass by 2--10$\times$ (with an average factor of $5.2\pm2.4$ increase)
from $z=2.3$ to now.  Therefore, it is likely that blobs are sites for
the formation of the brightest galaxies in what will become the typical
halos of rich clusters ($M_{\rm halo}$ $\sim$ 10$^{14}$\,$M_{\sun}$)
at the present epoch.

\subsection{Large-Scale Environment of Blobs}
\label{sec:lss}

The clustering of \lya\ blobs in the CDFS and the inferred large mass
of their individual halos imply that blobs inhabit overdense regions.
Here we test this hypothesis using the much larger population of
{\it compact} \lya\ emitters (LAEs) in the CDFS to trace large-scale
structure over tens of comoving Mpc and thus characterize the blob
environment independently.  \citet{Matsuda05} show that their blobs
lie near the intersection of large-scale filamentary structures in the
SSA22 overdensity.  \citet{Prescott08} use the surface number density
of LAEs to show that a giant blob, originally identified via its strong
MIR emission \citep{Dey05}, resides in a region $\sim 3\times$ more dense
than the edge of their survey field.  We apply counts-in-cell methodology
to the LAE spatial distribution in the CDFS to quantify the scale over
which any structure is coherent.  We then identify over-densities of
LAEs relative to their average number density in the field and compare
them to the spatial distribution of blobs\footnotemark.

\footnotetext{We do not detect any significant large-scale structure
in the other three survey fields.  Because survey depth and seeing vary
over the fields, it is not clear whether the non-detections arise from
a real absence of structure or poorer sensitivity.}

We select a sample of LAEs in the CDFS ($N$ $\sim$ 200 with {\sl NB} $<$
25.0, excluding the blobs) following the first step in our blob selection
procedure (\S\ref{sec:sample_selection}), but using a 2\arcsec-diameter
aperture to maximize the S/N of fainter point sources. In Table
\ref{tab:narrowband}, we list the 5$\sigma$ limiting magnitudes, which
are determined by measuring fluxes within the randomly placed apertures
in the sky background region.


\begin{figure}[!t]
\epsscale{1.15}
\plotone{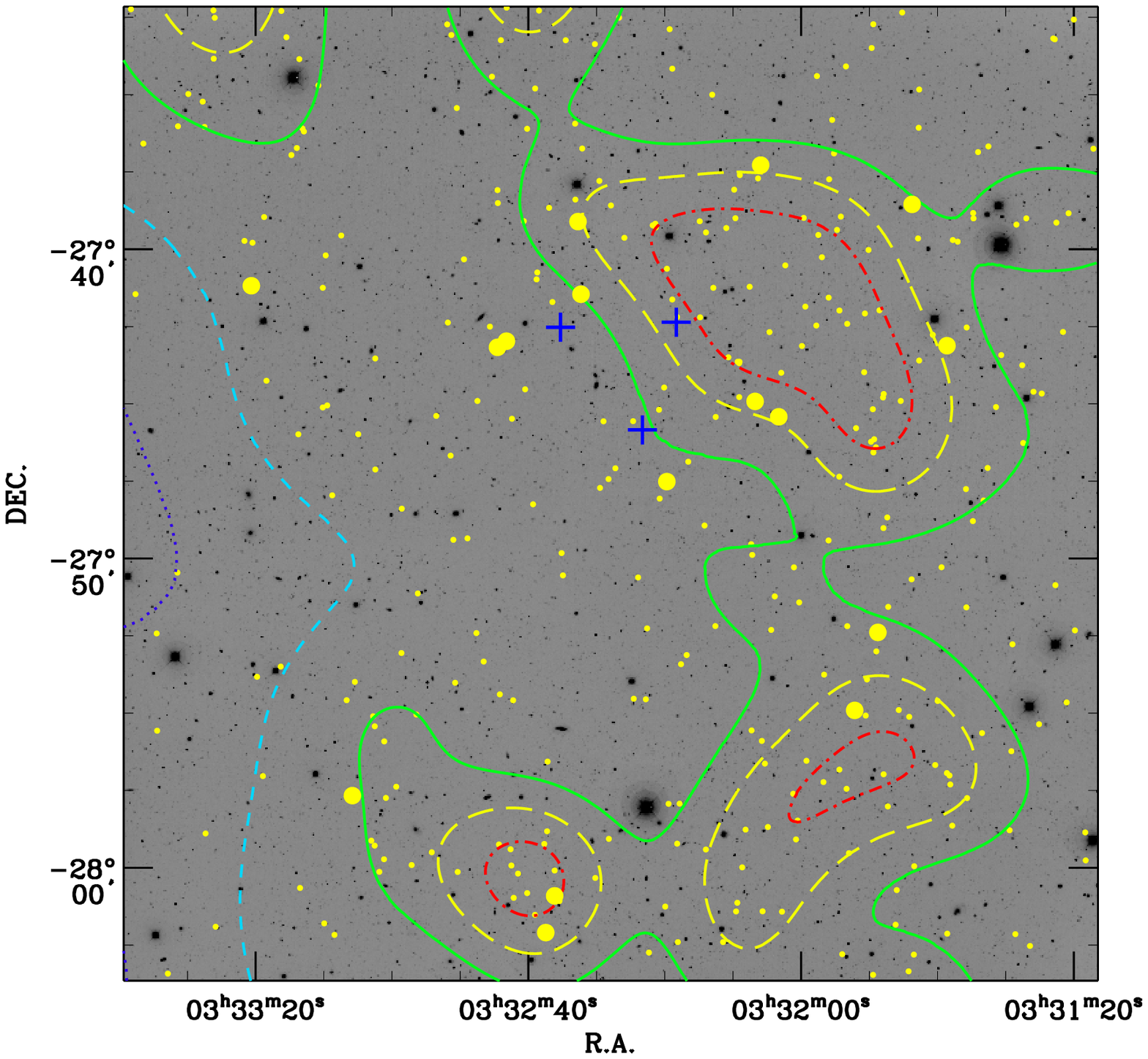}
\caption{
Surface density map ($\sim$30\arcmin$\times$30\arcmin =
50\,Mpc$\times$50\,Mpc) of compact \lya\ emitters in the CDFS.  The small
and big dots represent compact and extended \lya\ emitters (i.e., LAEs
and LABs), respectively.  The contours are spaced for $\delta_{\Sigma} =
({\Sigma} - \bar{\Sigma}) / \bar{\Sigma}$ = $-0.7$, $-0.5$, $0.0$, $0.5$,
$1.0$, where $\bar{\Sigma}$ is the the average LAE surface density of
the whole region.  The \lya\ blobs are located in the overdense region
traced by the LAEs. The three crosses represent overdensities identified
by \citet{Salimbeni09} using photometric redshifts ($z_{\rm phot} = 2.23,
2.28, 2.28$). A counts-in-cells analysis of the LAEs confirms coherent
large-scale structure over scales of at least $\sim$ 8 comoving Mpc.
}
\label{fig:surface_density}
\end{figure}


Using this LAE catalog, we test whether there is structure over various
spatial scales.  For each scale,  we count the number of LAEs within
circular cells of that scale radius randomly placed over the field.
Then we compare this counts-in-cells distribution with a Poisson
distribution using Kolmogorov-Smirnov (K-S) statistics to check if
this distribution deviates from Poisson noise. We repeat this test
with different cell sizes ranging from 1\arcmin\ to 8\arcmin. The
counts-in-cells (c-in-c) distributions deviate significantly (at $>
95.0\%$ confidence) from Poisson noise for cells larger than 5\arcmin,
indicating coherent large-scale structure over scales of at least $\sim$
8 comoving Mpc.

Figure \ref{fig:surface_density} shows the LAE distribution in the
CDFS field.  To estimate the overdensity of LAEs, we overlay the surface
overdensity $\delta_{\Sigma} = ({\Sigma} - \bar{\Sigma}) / \bar{\Sigma}$
contours on the map. Here $\bar{\Sigma}$ represents the average surface
density estimated from all the LAE candidates over the entire
field.  The overdensity maps are smoothed using the adaptive kernel
smoothing method developed by \citet{Ebeling06}. The FWHMs of the Gaussian
filters adopted by this algorithm ranges from 2.8\arcmin\ to 5.2\arcmin\
depending on the local surface density of LAEs within in the field.
We identify a belt-like large-scale structure elongated from north
to south.  The blob candidates are preferentially located within or at
the boundary of this structure.  Therefore, we conclude that the \lya\
blobs in the CDFS indeed trace overdensities in the early universe.

Recently, \citet{Salimbeni09} systematically searched for galaxy
overdensities within the GOODS-South field up to $z\sim2$ using
photometric redshifts that have a rms accuracy of  $\Delta z$/($1+z$)
= 0.03. In Figure \ref{fig:surface_density}, we show their three
overdensities (crosses) with redshifts similar to our survey ($z_{\rm
phot}$ $\simeq$ 2.23--2.28).  While the GOODS-South field does not
include the densest part of our surface density map, their overdensities
each include 19--23 members and are located near the boundary of the
LAE overdensity.  Spectroscopic confirmation of this system is required
to relate these systems with the structures revealed by our \lya\ blobs
and emitters.

Further studies are required to establish that the CDFS overdensity and
similar overdensities observed by other authors \cite[e.g.,][]{Prescott08}
are in fact ``proto-clusters.''  For now, we note that structure in the
CDFS at $z=2.3$ is coherent over at least 8 comoving Mpc and is likely
to contain tens of $\sim$ 10$^{13}$\,$M_{\sun}$ blob halos, or at least
$\sim$ 10$^{14}$\,$M_{\sun}$ worth of mass.  Were this overdensity to
grow typically and virialize by $z=0$, then its halo mass today would
be like that of the rich Coma cluster, i.e., $\sim$ 10$^{15}$\,$M_{\sun}$.

\clearpage
\section{Conclusions and Summary}
\label{sec:conclusion}

To understand what \lya\ blobs will become in the present-day universe
requires that we first constrain their number density, clustering,
and large-scale environment.  In order to obtain unbiased measures of
these quantities, we target four $\sim$30\arcmin$\times$30\arcmin\
fields, Chandra Deep Field South \cite[CDFS;][]{Brandt01},
Chandra Deep Field North \cite[CDFN;][]{Giacconi02}, and two COSMOS
\citep{Scoville07,Koekemoer07} subfields, with the NOAO Mayall and
Blanco-4m telescopes and a custom narrowband filter designed for \lya\
at $z\simeq2.3$.  The total area covered by our survey is 1.2\,deg$^2$.
Our sensitivity and selection criteria are comparable to that of the
largest previous blob survey \citep{Matsuda04}.

We discover 25 \lya\ blobs with \lya\ luminosities of \llya =
0.7--8 $\times$ $10^{43}$ \unitcgslum\ and isophotal areas of \Aiso =
10--60\,\sq\arcsec.  The transition from compact \lya\ emitters (LAEs;
\Aiso $\sim$\,a few \sq\arcsec) to the extended \lya\ blobs (\Aiso $>$
10\,\sq\arcsec) is continuous, suggesting that these two types of sources
are not distinct and that whatever mechanism or mechanisms power \lya\
blobs work over a wide range of luminosity and spatial extent.

Surprisingly, we find the majority of blobs (16/25) in one survey field,
the CDFS.  The six brightest (\llya\,$\gtrsim$ 1.5$\times$10$^{43}$
\unitcgslum) and largest (\Aiso $>$ 16\,\sq\arcsec) blobs are discovered
{\it only} in CDFS, indicating a strong field-to-field variation.
Using a simple analytic approximation for the underlying fluctuations
of the blob number density, we find that these large/bright blobs
have a field-to-field variance of $\sigma_v$ $\gtrsim$ 1.5 (150\%)
about their number density of $n$ $\sim$ $1.0^{+1.8}_{-0.6}$$\times$
$10^{-5}$ Mpc$^{-3}$.  This variance is large, significantly higher than
that of unresolved \lya\ emitters ($\sigma_v\sim$0.3 or 30\%).

To constrain the mass of the dark matter halo around each \lya\
blob, we compare the number density and clustering of blobs with the
counts-in-cells distribution of halos predicted from a 1 $h^{-1}$Gpc
cosmological N-body simulation.  At $z=2.3$, $n$ implies that
bright, large blobs could occupy halos of $M_{\rm halo}$ $\gtrsim$
$10^{13}$\,$M_{\sun}$ if most halos contain a detectable blob, i.e., the
detectability fraction is $\gtrsim 50\%$. The predicted variance in $n$
is consistent with that observed and corresponds to a bias of 7. Lower
detectability fractions (e.g., $\sim$ 10\%) require somewhat lower
halo masses ($\sim$ $5 \times 10^{12}$\,$M_{\sun}$) to reproduce $n$,
but the resulting variance is lower and further from the observed value.
Blob halos lie at the high end of the halo mass distribution at $z=2.3$
and are likely to evolve into the $\sim$10$^{14}$ $M_{\sun}$ halos
typical of galaxy clusters today.

The clustering and inferred halo mass of blobs suggest that they lie in
overdense environments.  The spatial distribution of LAEs confirms this
hypothesis:  a counts-in-cells analysis of the CDFS reveals coherent
large-scale structure over scales of at least $\sim$ 8 comoving Mpc
where both the LAEs and blobs cluster.

Given the strong field-to-field variance of \lya\ blobs, one must be
cautious in comparing blob number densities and luminosity functions
among different surveys.  We construct a reliable luminosity function
of \lya\ blobs from a deep, blind narrowband survey.
Larger volume blob surveys, combined with large volume
and/or higher resolution N-body simulations, will improve the constraints
on blob halo mass and detectability fraction, thus discriminating among
the possible energy sources of the extended \lya\ emission.

\acknowledgments

We thank an anonymous referee for helpful comments.
YY thanks the KPNO/CTIO staffs for observing support and the NOAO
graduate student travel fund.  We thank Yuichi Matsuda for providing
the narrowband images of his \lya\ blobs, Andrea Macci{\`o} for the
halo growth function data, and Masami Ouchi for his luminosity function
of LAEs.  We gratefully acknowledge the assistance of Toshihiko Kimura
at Asahi Spectra for his help in manufacturing our narrowband filter.
AIZ acknowledges support from the NSF Astronomy and Astrophysics Research
Program through grant AST-0908280 and from the NASA Astrophysics Data
Analysis Program through grant NNX10AD47G.

\smallskip
Facilities: \facility{Mayall (MOSAIC I), Blanco (MOSAIC II)}

\clearpage


\newcommand\ff[1]{\tablenotemark{#1}}
\begin{deluxetable}{cccccccccc}
\tablewidth{0pt}
\tabletypesize{\small}
\tabletypesize{\scriptsize}
\tablecaption{Narrowband Observations}
\tablehead{
\colhead{Field}&
\colhead{R.A.}&
\colhead{Dec.}&
\colhead{Observing Date}&
\colhead{Telescope}&
\colhead{Exposure}&
\colhead{Depth\tablenotemark{a}}&
\colhead{Survey Area}&
\colhead{Seeing}&
\colhead{Pixel Scale\tablenotemark{b}} \\ 
\colhead{}&         
\colhead{(J2000)}&  
\colhead{(J2000)}&  
\colhead{}&         
\colhead{}&         
\colhead{(hour)}&   
\colhead{(AB mag)}& 
\colhead{}&         
\colhead{}&         
\colhead{}          
}
\startdata
  CDF-S        & 03:32:27.8  &      -27:47:56 & Nov. 2007        & Blanco 4m &    10    &  25.65 & 31\farcm6 $\times$ 31\farcm6 & 1.0\arcsec &       0\farcs27   \\ 
  CDF-N        & 12:36:50.9  &\phm{-}62:11:48 & May\phm{.}  2007 & Mayall 4m &    10    &  25.27 & 29\farcm5 $\times$ 29\farcm5 & 1.3\arcsec &       0\farcs30   \\ 
  COSMOS1      & 09:59:16.9  &\phm{-}01:55:19 & Feb. 2009        & Blanco 4m &    7.7   &  25.27 & 36\farcm0 $\times$ 36\farcm3 & 1.2\arcsec &       0\farcs27   \\ 
  COSMOS2      & 09:59:16.8  &\phm{-}02:31:19 & Feb. 2009        & Blanco 4m &    7.2   &  25.25 & 36\farcm0 $\times$ 36\farcm3 & 1.2\arcsec &       0\farcs27   \\
\enddata
\tablenotetext{a}{5$\sigma$ detection limit for 2$\arcsec$-diameter aperture.}
\tablenotetext{b}{Pixel scales of final combined images, which are determined by the largest pixel scale between narrowband and broadband images.}
\label{tab:narrowband}
\end{deluxetable}




\begin{deluxetable}{cccccccc}
\tablewidth{0pt}
\tabletypesize{\small}
\tabletypesize{\scriptsize}
\tablecaption{Summary of Broadband Images}
\tablehead{
\colhead{Field                   }&
\colhead{Band                    }&
\colhead{Effective Wavelength    }&
\colhead{Band Width              }&
\colhead{Depth                   }&
\colhead{Seeing                  }&
\colhead{Telescope and Instrument}&
\colhead{Reference               }\\
\colhead{        }&   
\colhead{        }&   
\colhead{(\AA)   }&   
\colhead{(\AA)   }&   
\colhead{(AB mag)}&   
\colhead{(arcsec)}&   
\colhead{        }&   
\colhead{        }    
}
\startdata
  CDF-S  & {\sl U} &         3505         &     626     &    26.0\tablenotemark{a}      &   1.05   & ESO 2.2m WFI              &  \citet{Gawiser06b} \\
         & {\sl B} &         4600         &     915     &    26.9\tablenotemark{\phn}   &   0.95   & ESO 2.2m WFI              &                     \\ 
  CDF-N  & {\sl U} &         3648         &     387     &    27.1\tablenotemark{b}      &   1.26   & KPNO MOSAIC I             &  \citet{Capak04}    \\
         & {\sl B} &         4428         &     622     &    26.9\tablenotemark{\phn}   &   0.71   & Subaru Suprime-Cam        &                     \\
  COSMOS & {\sl U} &         3798         &     720     &    26.4\tablenotemark{b}      &   0.90   & CFHT Megaprime            &  \citet{Capak07}    \\
         & {\sl B} &         4460         &     897     &    27.3\tablenotemark{\phn}   &   0.95   & Subaru Suprime-Cam        &                     
\enddata
\tablenotetext{a}{CDF-S:  5$\sigma$ detection limit corrected for infinite aperture.}
\tablenotetext{b}{CDF-N and COSMOS  5$\sigma$ detection limit for 3\arcsec diameter aperture.}
\label{tab:broadband}
\end{deluxetable}




\begin{deluxetable}{lcccccc}
\tablecolumns{7} 
\tablewidth{0pt}
\tabletypesize{\small}
\tabletypesize{\scriptsize}
\tablecaption{Properties of L$\lowercase{\rm y}$$\alpha$ Blob Candidates}
\tablehead{
\colhead{ID}& 
\colhead{R.A.}&
\colhead{Dec.}&
\colhead{$L_{\rm tot}$}&
\colhead{$L_{\rm iso}$}&
\colhead{Area}&
\colhead{Note}\\
\colhead{}& 
\colhead{(J2000)}&
\colhead{(J2000)}&
\colhead{($10^{43}$\unitcgslum)}&
\colhead{($10^{43}$\unitcgslum)}&
\colhead{(\sq\arcsec)}&
\colhead{}
\\ \cline{1-7}\\
\multicolumn{7}{c}{Extended Chandra Deep Field South}
}
\startdata
      CDFS-LAB01 &     03 32 36.1 &    $-$28 00 54.5 &     7.81 $\pm$  0.28 &       8.00 &     61.6 $\pm$   3.7 & confirmed \\  
      CDFS-LAB02 &     03 33 20.6 &    $-$27 41 10.8 &     3.26 $\pm$  0.17 &       2.88 &     37.9 $\pm$   3.3 & confirmed \\ 
      CDFS-LAB03 &     03 31 52.1 &    $-$27 54 54.6 &     3.23 $\pm$  0.29 &       2.72 &     43.2 $\pm$   4.4 & \\ 
      CDFS-LAB04 &     03 32 05.9 &    $-$27 37 16.7 &     2.64 $\pm$  0.15 &       2.40 &     30.9 $\pm$   2.1 & confirmed \\ 
      CDFS-LAB05 &     03 31 48.7 &    $-$27 52 23.3 &     1.72 $\pm$  0.39 &       1.14 &     21.3 $\pm$   6.2 & \\ 
      CDFS-LAB06 &     03 32 19.6 &    $-$27 47 30.8 &     1.53 $\pm$  0.11 &       1.57 &     16.5 $\pm$   1.8 & \\ 
      CDFS-LAB07 &     03 32 03.2 &    $-$27 45 25.0 &     1.48 $\pm$  0.17 &       1.56 &     18.5 $\pm$   2.5 & \\ 
      CDFS-LAB08 &     03 31 43.7 &    $-$27 38 32.9 &     1.21 $\pm$  0.22 &       1.32 &     19.3 $\pm$   4.5 & \\ 
      CDFS-LAB09 &     03 32 44.5 &    $-$27 43 10.2 &     1.17 $\pm$  0.15 &       0.87 &     14.3 $\pm$   2.0 & \\ 
      CDFS-LAB10 &     03 32 37.4 &    $-$28 02 05.7 &     1.14 $\pm$  0.25 &       0.71 &     12.6 $\pm$   2.6 & confirmed \\ 
      CDFS-LAB11 &     03 32 43.2 &    $-$27 42 58.3 &     1.09 $\pm$  0.08 &       1.02 &     10.3 $\pm$   1.3 & \\ 
      CDFS-LAB12 &     03 32 06.7 &    $-$27 44 55.2 &     1.08 $\pm$  0.16 &       0.73 &     13.3 $\pm$   2.4 & \\ 
      CDFS-LAB13 &     03 32 32.7 &    $-$27 39 06.3 &     1.03 $\pm$  0.09 &       0.94 &     12.0 $\pm$   1.5 & \\ 
      CDFS-LAB14 &     03 32 32.2 &    $-$27 41 27.2 &     0.98 $\pm$  0.09 &       0.93 &     12.8 $\pm$   1.5 & confirmed \\ 
      CDFS-LAB15 &     03 31 38.6 &    $-$27 43 07.0 &     0.96 $\pm$  0.08 &       0.81 &     11.8 $\pm$   1.1 & \\ 
      CDFS-LAB16 &     03 33 05.8 &    $-$27 57 40.0 &     0.69 $\pm$  0.18 &       0.62 &     10.3 $\pm$   3.0 & \\ 
\cutinhead{Chanda Deep Field North} 
      CDFN-LAB01 &     12 35 35.2 &    $+$62 14 28.5 &     1.31 $\pm$  0.11 &       1.30 &     13.2 $\pm$   1.8 & \\ 
      CDFN-LAB02 &     12 35 30.2 &    $+$62 01 39.2 &     1.15 $\pm$  0.12 &       0.94 &     11.7 $\pm$   1.7 & marginal? \\ 
      CDFN-LAB03 &     12 36 09.9 &    $+$61 57 16.6 &     1.01 $\pm$  0.12 &       0.94 &     12.2 $\pm$   1.9 & marginal? \\ 
      CDFN-LAB04 &     12 36 59.2 &    $+$62 24 35.2 &     0.99 $\pm$  0.15 &       0.80 &     13.0 $\pm$   2.0 & \\ 
      CDFN-LAB05 &     12 38 18.4 &    $+$62 04 04.1 &     0.72 $\pm$  0.13 &       0.61 &     11.1 $\pm$   1.9 & \\ 
\cutinhead{Cosmic Origins Evolution Survey Field} 
    COSMOS-LAB01 &     09 59 23.9 &    $+$01 55 11.7 &     1.24 $\pm$  0.19 &       0.87 &     14.6 $\pm$   5.2 & marginal? \\ 
    COSMOS-LAB02 &     09 59 14.2 &    $+$01 48 43.1 &     0.92 $\pm$  0.19 &       1.05 &     11.9 $\pm$   2.3 & marginal? \\ 
    COSMOS-LAB03 &     09 58 12.8 &    $+$01 52 56.2 &     0.84 $\pm$  0.16 &       0.70 &     12.5 $\pm$   2.5 & \\ 
    COSMOS-LAB04 &     09 58 49.6 &    $+$02 30 49.9 &     1.01 $\pm$  0.13 &       0.92 &     11.8 $\pm$   2.1 & 
\enddata
\label{tab:properties}
\tablecomments{
The note column indicates whether the redshifts of the candidates are
spectroscopically confirmed with the \lya\ and/or H$\alpha$ line and
whether they are marginally resolved (labelled ``marginal'').
}
\end{deluxetable}




\begin{deluxetable}{lcccc}
\tablewidth{0pt}
\tabletypesize{\small}
\tabletypesize{\scriptsize}
\tablecaption{Blob Statistics}
\tablehead{
\colhead{}&
\colhead{CDFS   }&
\colhead{CDFN   }&
\colhead{COSMOS1}&
\colhead{COSMOS2}
}
\startdata  
 $A_{\rm iso}$ $>$ 16\,\sq\arcsec                     &       8   &      0   &       0   &        0 \\
 $A_{\rm iso}$ $=$ 10--16\,\sq\arcsec                 &       8   &      5   &       3   &        1 \\
\tableline 
 $N_{\rm blob}$(bright/large sample)\tablenotemark{a} &       6   &      0   &       0   &        0 \\
 $N_{\rm blob}$(entire sample)                        &      14   &      5   &       3   &        1 \\
 Effective area [deg$^2$]                             &   0.270   &  0.241   &   0.352   &    0.362 
\enddata
\label{tab:stat}
\tablenotetext{a}{$N_{\rm blob}$ is the number of blobs corrected to the
same depth and seeing for all four survey fields. We use $N_{\rm blob}$
for estimating the number density, its variance, and the blob halo mass
in \S\ref{sec:cosmic_variance_est}.
\tablecomments{The bright, large blob sample consists of blobs with
\llya\,$\gtrsim$ 1.5$\times$10$^{43}$ \unitcgslum\ and $A_{\rm iso}$
$>$ 16\,\sq\arcsec.}
}
\end{deluxetable}




\begin{deluxetable}{cccccccc}
\tablecolumns{8}
\tablewidth{0pt}
\tabletypesize{\small}
\tabletypesize{\scriptsize}
\tablecaption{Field-to-Field Variance and Number Density of Blobs}
\tablehead{
\colhead{     }&
\multicolumn{3}{c}{Variance $\sigma_v$}& 
\colhead{     }&
\multicolumn{3}{c}{Number per 0.25 deg$^2$ $\bar{N}$} \\
\cline{2-4} \cline{6-8} \\
\colhead{    }&
\colhead{Min }&
\colhead{Best}&
\colhead{Max }&
\colhead{    }&
\colhead{Min }&
\colhead{Best}&
\colhead{Max }\\
\cline{1-8}\\
\multicolumn{8}{c}{Bright, Large Blobs with \Aiso $>$ 16\,\sq\arcsec}
}
\startdata
    Joint     &  1.49     (0.58)   &   \nodata &  \nodata  &  &     0.27 \phm{(.000)} & \nodata &  \nodata           \\
    Marginal  &  3.11     (0.99)   &   \nodata &  \nodata  &  &     0.40     (0.17)   &   1.15  &   3.27      (8.30) \\

\cutinhead{Entire Sample with \Aiso $>$ 10\,\sq\arcsec} 

    Joint     &  0.26     (0.10)   &     0.87  &    5.21   &  &     2.27     (1.56)   &   4.76  &   21.7 \phm{(0.00)}\\
    Marginal  &  0.57     (0.31)   &     1.20  &    3.51   &  &     2.91     (2.13)   &   4.76  &   10.3      (27.1) \\

\enddata
\label{tab:cosmic_variance}
\tablecomments{
The number density $\bar{N}$ represents the number of blobs per 0.25
deg$^2$. In the parentheses, we list the 2$\sigma$ limits.
}
\end{deluxetable}




\begin{deluxetable}{c c ccc}
\tablecolumns{10} 
\tablewidth{0pt}
\tabletypesize{\small}
\tabletypesize{\scriptsize}
\tablecaption{cumulative Luminosity Function of Ly$\alpha$ Blobs}
\tablehead{
\colhead{Log($L_{\rm Ly\alpha}$)}&   
\colhead{\smallskip}&   
\multicolumn{3}{c}{$n(>L)$ [$\times10^{-5}$\,Mpc$^{-3}]$} \\
\cline{3-5}\\
\colhead{}&   
\colhead{}&   
\colhead{CDFS\tablenotemark{a}}&   
\colhead{All\tablenotemark{b}}&  
\colhead{SSA22\tablenotemark{c}}
}
\startdata
 43.90 & & \nodata                 & \nodata                 &  $  \X1.54 \pm     1.09$ \\ 
 43.70 & & $  \X0.83 \pm     0.83$ & $    0.18 \pm     0.18$ &  $  \X2.31 \pm     1.33$ \\
 43.50 & & $  \X2.50 \pm     1.44$ & $    0.55 \pm     0.32$ &  $  \X3.08 \pm     1.54$ \\
 43.30 & & $  \X3.33 \pm     1.67$ & $    0.73 \pm     0.37$ &  $  \X5.38 \pm     2.04$ \\
 43.10 & & $  \X5.83 \pm     2.20$ & $    1.47 \pm     0.52$ &  $   10.77 \pm     2.88$ \\
 42.90 & & $   12.49 \pm     3.22$ & $    4.23 \pm     0.88$ &  $   18.46 \pm     3.77$ \\
 42.70 & & $   13.32 \pm     3.33$ & $    4.59 \pm     0.92$ &  $   26.92 \pm     4.55$ \\
\enddata
\label{tab:luminosity_function}
\tablenotetext{a}{LF from blobs in CDFS.}

\tablenotetext{b}{LF from all four survey fields (CDFS, CDFN, COSMOS1, and COSMOS2).}

\tablenotetext{c}{We also list the LF from \citet{Matsuda04} for comparison.}
\end{deluxetable}


\clearpage


\begin{deluxetable}{rccccccccc}
\tablecolumns{10} 
\tablewidth{0pt}
\tabletypesize{\small}
\tabletypesize{\scriptsize}
\tablecaption{Counts-in-cells Distribution for Bright, Large Blobs}
\tablehead{
\colhead{}&   
\colhead{}&   
\colhead{}&   
\multicolumn{3}{c}{N-Body Simulation}&
\colhead{}&   
\multicolumn{3}{c}{Poisson Distribution}\\
\cline{4-6} \cline{8-10}\\
\colhead{$f_{\rm D}$}&   
\colhead{$M_{\rm min} (M_{\sun})$}&   
\colhead{$\sigma_v$}&  
\colhead{$P(N=0)$}& 
\colhead{$P(N\geqslant6)$}& 
\colhead{Probability}&
\colhead{}&
\colhead{$P(N=0)$}& 
\colhead{$P(N\geqslant6)$}& 
\colhead{Probability}\\
\colhead{(1)}& 
\colhead{(2)}&
\colhead{(3)}&
\colhead{(4)}&
\colhead{(5)}&
\colhead{(6)}&
\colhead{}&
\colhead{(7)}&
\colhead{(8)}&
\colhead{(9)}
}
\startdata
       100\%  &   1.66$\times10^{13}$  &      1.03  & 41.5\%  &   4.40\%    &          1.25\% & &    23.3\% &  0.440\% &     0.022\%  \\
        50\%  &   1.16$\times10^{13}$  &      0.93  & 38.5\%  &   3.65\%    &          0.83\% & &    22.9\% &  0.370\% &     0.018\%  \\
        25\%  &   7.97$\times10^{12}$  &      0.83  & 35.8\%  &   3.27\%    &          0.60\% & &    23.4\% &  0.300\% &     0.015\%  \\
      12.5\%  &   5.32$\times10^{12}$  &      0.76  & 35.6\%  &   2.51\%    &          0.45\% & &    23.8\% &  0.370\% &     0.020\%  \\
\enddata
\label{tab:duty_cycle}
\tablecomments{
(1) detectability fraction,
(2) dark matter halo mass,
(3) variance [Eq. (\ref{eq:variance})],
(4) probability of non-detection,
(5) probability of finding 6 or more blobs,
(6) posterior probability of finding 6 or more blobs in one field and none in the three other fields,
(7)--(9) same as (4)--(6) but for Poisson distribution.
}
\end{deluxetable}


\appendix

\section{Comparison with Matsuda et al.\ L\lowercase{y$\alpha$} Blob Sample }
\label{sec:blob_comparison}

It is difficult to compare the properties of \lya\ blob samples
among different surveys because of non-uniform selection criteria and
different imaging depth. Here we compare our $z=2.3$ sample to that of
\citet{Matsuda04} at $z=3.1$. We repeat the procedures for the recovery
test described in \S\ref{sec:sample_selection} \cite[see also][]{Yang09},
pasting thumbnail images provided by \citet{Matsuda04} into our CDFS
image. Because the narrowband filter bandwidths of the two surveys are
similar, we do not make any correction for the difference in filter
transmission, but we scale the apparent size and the surface brightness
accounting for the different redshifts.

\begin{figure}
\epsscale{1.0}
\vspace{1cm}
\plottwo{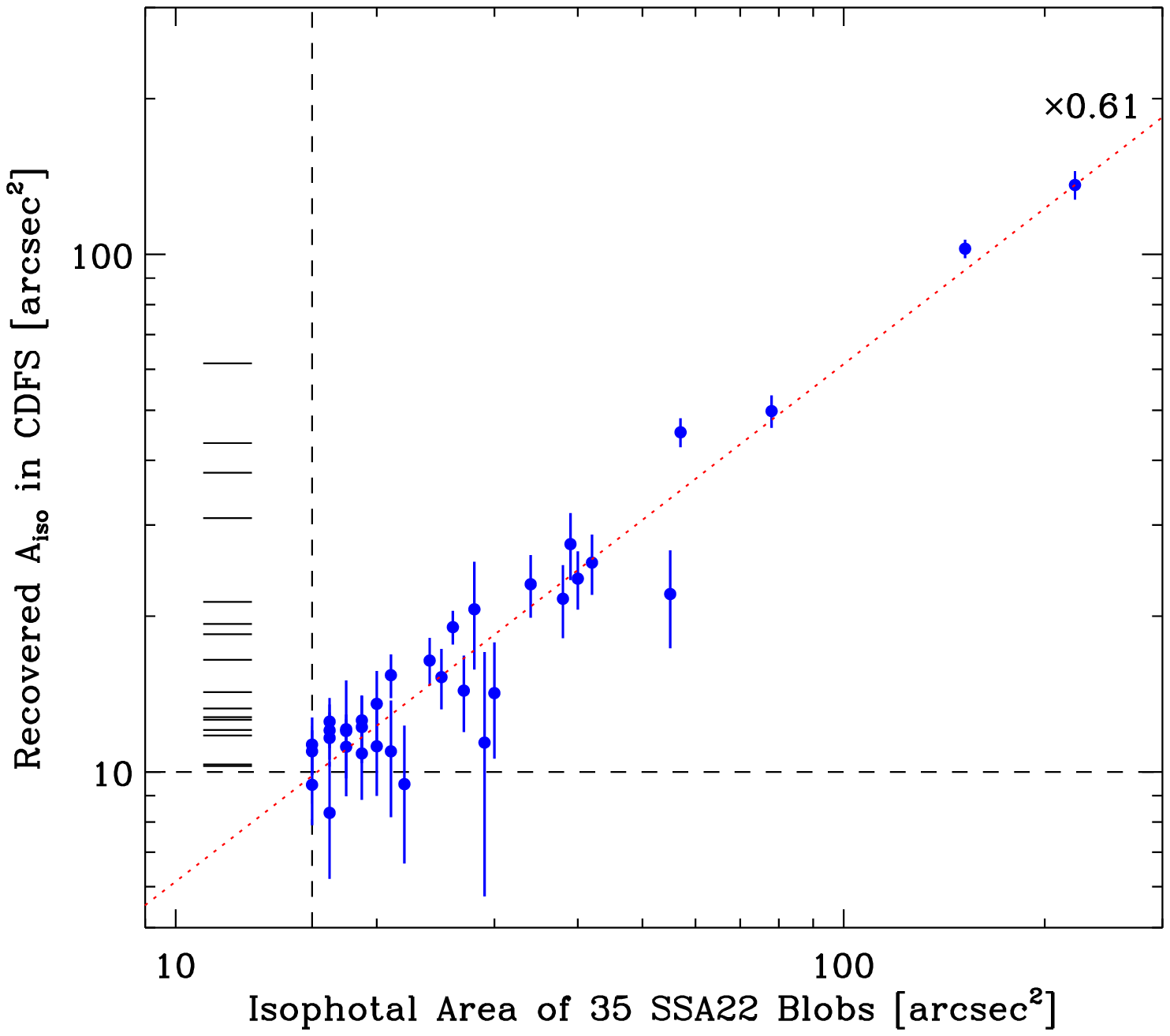}
        {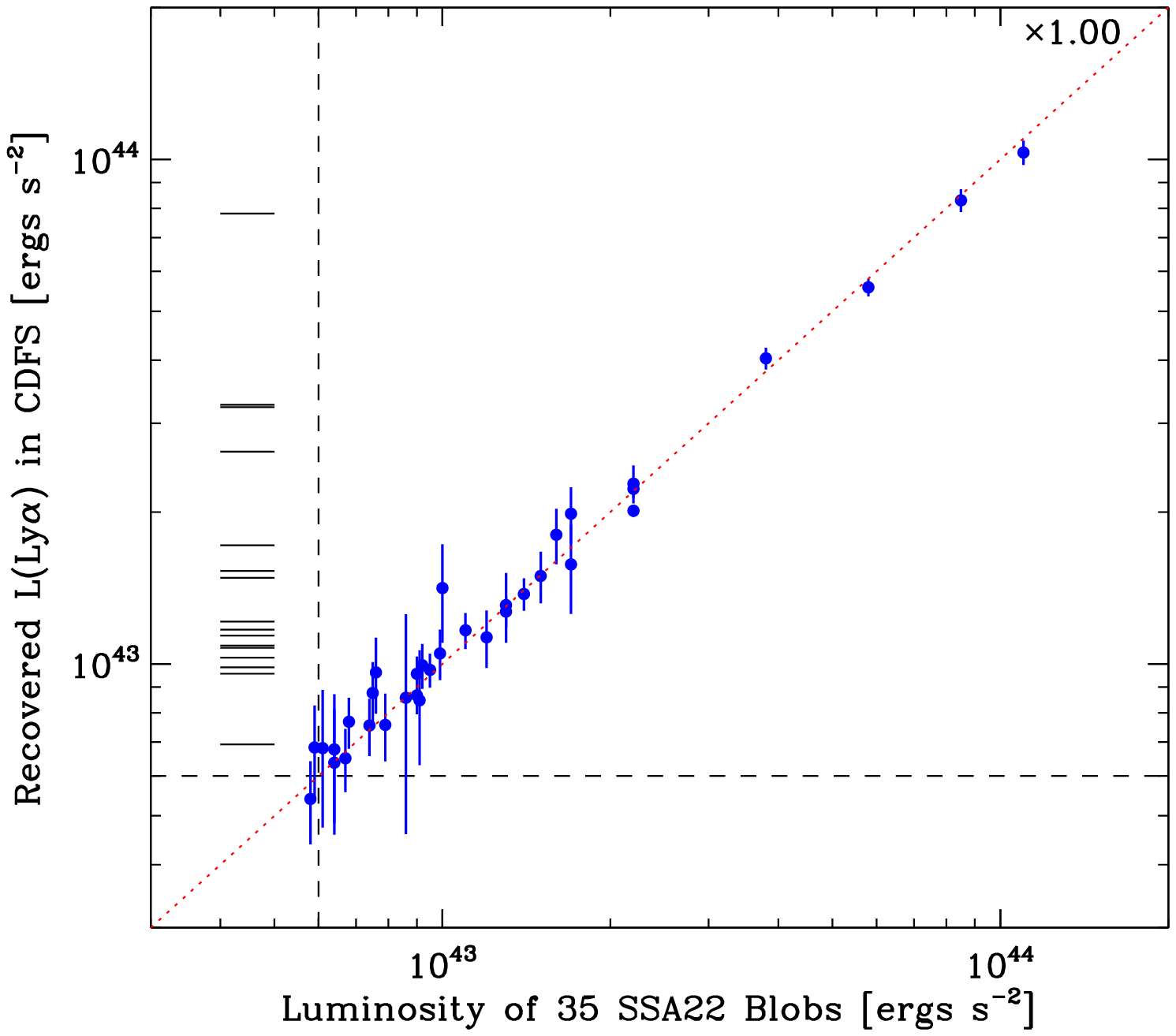}
\caption{
Comparison in \llya\ and \Aiso between our 16 CDFS blobs (horizontal bars)
and the 35 \cite{Matsuda04} blobs in the SSA22 field (filled circles).
The vertical and horizontal dashed lines indicate the size selection
criteria for the extended sources adopted by \citet{Matsuda04} and by
our work, respectively.  The dotted lines represent the average ratios
($\sim$0.61 and 1.0) between originally reported and the recovered
values in \Aiso and \llya.  The recovery test of the \cite{Matsuda04}
blobs when pasted into our CDFS field shows that we can detect all blobs
like theirs in our survey.
}
\label{fig:comparison_blob}
\end{figure}

Figure \ref{fig:comparison_blob} shows the \lya\ luminosity and isophotal
area recovered from this test as a function of the input \llya\ and \Aiso
for the 35 \citet{Matsuda04} blobs.  Due to the slightly shallower depth
of our survey, the \citeauthor{Matsuda04} blobs would look smaller by a
factor of 61\% than the originally reported sizes.  Note that one expects
a 86\% decrease in area purely from the differences in angular diameter
distance between two survey redshifts.  The vertical and horizontal
dashed lines indicate the size selection criteria for the extended
sources adopted by \citet{Matsuda04} and by our work, respectively.
Most \citet{Matsuda04} blobs are recovered as larger than 10\,\sq\arcsec,
confirming the capability of our survey for detecting them.  The line
luminosities are also recovered well, so there is no bias in the
luminosity measurements.

\end{document}